\documentclass[aps,prd,twocolumn,floatfix,preprintnumbers,nofootinbib, superscriptaddress]{revtex4-1}

\usepackage{blindtext}
\usepackage{epsf,graphicx,xcolor}
\usepackage{amsmath,amssymb,amstext}
\allowdisplaybreaks

\usepackage{mathtools}
\newcommand{\sla}[1]{\ensuremath{\mathrlap{\!\not{\phantom{#1}}}#1}}
\usepackage{epstopdf}
\usepackage{url,hyperref}
\usepackage[utf8]{inputenc}
\usepackage[english]{babel}
\usepackage{float}

\newcommand{\pvec}[1]{\vec{#1}\mkern2mu\vphantom{#1}}

\begin{document}

\title{Leading-order QED radiative corrections to timelike Compton scattering on the proton}
\author{Matthias Heller}
\affiliation{Institut f\"ur Kernphysik and $\text{PRISMA}^+$ Cluster of Excellence, Johannes Gutenberg Universit\"at, D-55099 Mainz, Germany}
\author{Niklas Keil}
\affiliation{Institut f\"ur Kernphysik and $\text{PRISMA}^+$ Cluster of Excellence, Johannes Gutenberg Universit\"at, D-55099 Mainz, Germany}
\author{Marc Vanderhaeghen}
\affiliation{Institut f\"ur Kernphysik and $\text{PRISMA}^+$ Cluster of Excellence, Johannes Gutenberg Universit\"at, D-55099 Mainz, Germany}

\date{\today}

\begin{abstract}

We evaluate the leading-order QED radiative corrections to the timelike Compton scattering (TCS) process $\gamma p \to l^- l^+ p$. We study these corrections in two energy regimes using different models for the TCS amplitude. In the low-energy regime we calculate the contribution due to the proton and its lowest-energy excitation, the $\Delta(1232)$ resonance. In the high-energy near-forward kinematical regime we calculate the TCS amplitude in a handbag approach in terms of Generalized Parton Distributions (GPDs). On the level of cross sections we find the QED radiative corrections to be in the $5 -10\%$ range in the low-energy regime and around $20\%$ in the high-energy regime. We show that in both the di-lepton forward-backward asymmetry as well as in the photon beam helicity asymmetry these corrections nearly cancel out, making them gold-plated observables to extract the real and imaginary parts of the TCS amplitude. We demonstrate in particular the sensitivity of these asymmetries on GPD parameterizations for a recent CLAS12@JLab TCS experiment.  
\end{abstract}

\maketitle

\section{Introduction}

The virtual Compton scattering (VCS) process is a versatile tool to unravel the proton electromagnetic structure beyond the information contained in its elastic form factors. The $e^- p \to e^- p \gamma$ reaction which accesses the virtual Compton scattering process with an incoming photon with spacelike virtuality and outgoing real photon has been studied extensively both at low and high energies, see Ref.~\cite{Guichon:1998xv} for an early review. At low energies, it allows to access generalized polarizabilities of the proton~\cite{Guichon:1995pu}, which encode the spatial deformations of the quark charge densities in a proton upon applying an external electromagnetic field. 
These observables have been extracted over the past two decades at the electron scattering facilities MIT-Bates, MAMI, and Jefferson Lab (JLab), see Ref.~\cite{Fonvieille:2019eyf} for a recent review.  At high energies and for near-forward kinematics, the $e^- p \to e^- p \gamma$ process is closely related to deep-inelastic scattering. In this regime, pertubative Quantum Chromo Dynamics (QCD) allows to express the proton structure entering the deeply virtual Compton scattering (DVCS) process through Generalized Parton Distributions (GPDs), which access the correlation between the longitudinal momentum distribution of partons in a proton and their two-dimensional transverse spatial distributions. We refer the reader to Refs.~\cite{Ji:1996ek,Mueller:1998fv,Radyushkin:1996nd,Ji:1996nm}
for the original articles on GPDs and to
Refs.~\cite{Goeke:2001tz,Diehl:2003ny,Belitsky:2005qn,Boffi:2007yc,Guidal:2013rya,Kumericki:2016ehc} for reviews of the field.   
Accessing the resulting three-dimensional momentum-spatial  distributions of valence quarks in a nucleon through exclusive processes has been one of the driving motivations for the JLab 12 GeV upgrade~\cite{osti_1345054}. Furthermore, accessing the 
sea-quark and gluonic structure of nucleons and nuclei through such processes is one of the main science questions that will be addressed at the future Electron-Ion Collider (EIC) machine~\cite{Accardi:2012qut}. 

The information accessed in the $e^- p \to e^- p \gamma$ reaction can be complemented through the $\gamma p \to e^-e^+ p$ reaction which accesses the timelike Compton scattering (TCS) process with incoming real photon and outgoing timelike photon, through production of a di-lepton pair. In the near-forward kinematical regime in which the timelike photon has a large virtuality, the non-perturbative information entering the TCS amplitude can also be expressed in terms of GPDs~\cite{Berger:2001xd}. Furthermore, a combination of both DVCS and TCS observables allows for a stringent test of the applicability of the underlying QCD factorization theorem at these kinematics. 
Such measurement of the TCS process at large timelike virtuality has been proposed by CLAS12@JLab~\cite{JLabtcs:2020}, and recently first data of this experiment have been reported~\cite{Chatagnon:2020egu,Chatagnon:2020}. 

A further extension of either the DVCS or TCS process in the high-energy near-forward region has been proposed through the $e^- p \to e^- p l^-l^+$ reaction (with $l^-$ either an $e^-$ or $\mu^-$), which accesses the double deeply virtual Compton scattering (DDVCS) process with incoming spacelike photon and outgoing timelike photon. The DDVCS process is of particular interest as it allows to extend the beam spin asymmetry measurements of GPDs into the ERBL domain~\cite{Guidal:2002kt,Belitsky:2002tf}.
Recently a feasibility study of the DDVCS experiment has shown  that the SoLID@JLab project with its high-luminosity and large acceptance is very promising to perform such measurements~\cite{Accardi:2020swt}. 

The measurement of the $e^- p \to e^- p l^-l^+$ process has recently also been proposed in the low-energy region in order to help reducing the theoretical model error in estimates of the hadronic correction to the muonic hydrogen Lamb shift. Over the past decade, measurements of the 2S-2P  
Lamb shift in muonic hydrogen ($\mu H$) have reported a proton charge radius with an order of magnitude improvement in its precision~\cite{Pohl:2010zza,Antognini:1900ns} as compared to the precision obtained in electron scattering. Initially these muonic atom extractions of the proton charge radius disagreed by around 5.6 standard deviations with the values obtained from energy level shifts in electronic hydrogen~\cite{Mohr:2015ccw} or from electron-proton scattering~\cite{Bernauer:2010wm,Bernauer:2013tpr}, which triggered a lot of activity in recent years, see Refs.~\cite{Carlson:2015jba,Pohl:2013yb} for some reviews.  
A new Lamb shift measurement in electronic hydrogen~\cite{Bezginov:2019mdi} as well as a new experiment using electron scattering~\cite{Xiong:2019umf} are now both in support of the smaller proton radius value obtained by muonic measurements. To fully clarify the situation, further experiments with electron beams at MAMI and MESA~\cite{Denig:2016tpq}, with muon beams at the Paul Scherer Institute \cite{Gilman:2017hdr} and at CERN \cite{Denisov:2018unj}, or through a direct comparison of $\gamma p \to e^-e^+ p$ versus $\gamma p \to \mu^- \mu^+ p$~\cite{Pauk:2015oaa} have been proposed or are underway. 
All these experiments aim at extracting the proton charge radius with improved precision. Experimentally so far, the most precise measurement is coming from the $\mu H$ Lamb shift measurements, for which the proton form factors and polarizabilities are required as theoretical input to estimate the next-order proton structure corrections. Those are at present the theoretical limitation on the proton radius extraction from Lamb shift measurements. It has been demonstrated however how these proton structure corrections can be empirically constrained through  measurements of the forward-backward asymmetry in the $e^- p\rightarrow e^- p e^+e^-$ process~\cite{Pauk:2020gjv}. 

To extract all of the above proton structure information from the single or double virtual Compton scattering process requires an estimate of the Quantum-Electrodynamic (QED) radiative corrections to these processes. 
For the $e^- p \to e^- p \gamma$ reaction, a detailed study of the radiative corrections has been performed~\cite{Vanderhaeghen:2000ws} and has been applied to existing data~\cite{Fonvieille:2019eyf}. 
In the present work, we report on the calculation of the leading-order QED corrections to the reaction $\gamma p \rightarrow e^-e^+ p$ in two different energy regimes, generalizing our calculation of Ref.~\cite{Heller:2018ypa,Heller:2019dyv} to include the TCS process. We study the radiative corrections on the cross section as well as the di-lepton forward-backward asymmetry and the photon beam helicity asymmetry. The present work will set the stage for a future study of the leading-order QED radiative corrections to the double VCS reaction $e^- p \to e^- p l^-l^+$. 

The outline of the present paper is as follows. In Section~\ref{sec:tree_level} we introduce the contributing Bethe-Heitler (BH) and TCS amplitudes to the $\gamma p \to l^-l^+ p$ reaction at tree level and define the relevant kinematic variables. In Section~\ref{sec:compton} we describe two different models for the double virtual Compton amplitude which are tailored for applications in two different energy regimes. In the low-energy regime, motivated for applications to describe the hadronic structure in precision muonic atom spectroscopy, we calculate the contribution due to the proton and its lowest-energy excitation, the $\Delta$(1232) resonance. In the high-energy near-forward kinematical regime in which at least one of the photons has a large virtuality, we use a QCD factorization theorem to describe the double virtual Compton amplitude on the proton in terms of a Compton amplitude on the quark convoluted with the non-perturbative structure of the proton encoded in the GPDs. In Sections~\ref{sec:vac-pol}, \ref{sec:one_loop_dvcs} and \ref{sec:one_loop_bh} we calculate the virtual one-loop QED corrections due to vacuum polarization, as well as due to photons attached to the leptonic lines of the BH and TCS amplitudes. In both cases we calculate the corrections on the level of the amplitude, allowing for the calculation of polarized cross sections. In Section~\ref{sec:soft_real} we calculate the soft-photon emission contributing to the process in which the di-lepton pair is measured. Including the soft-photon radiation gives infrared finite results for the observables.  In Section~\ref{sec:results}, we present our numerical results, and show the effect of the radiative corrections on the cross sections as well as on the forward-backward and photon beam helicity asymmetries. We show results both in the $\Delta(1232)$ resonance region as well as in the kinematical regime of the CLAS12@JLab TCS experiment. For the latter, we show the sensitivity of the cross section and asymmetries on the underlying GPD parameterization. We conclude in Section~\ref{sec:conclusion}.

\section{Bethe-Heitler and timelike Compton Scattering processes at tree level}
\label{sec:tree_level}
\begin{figure}[h]
	\centering
  \includegraphics[scale=0.6]{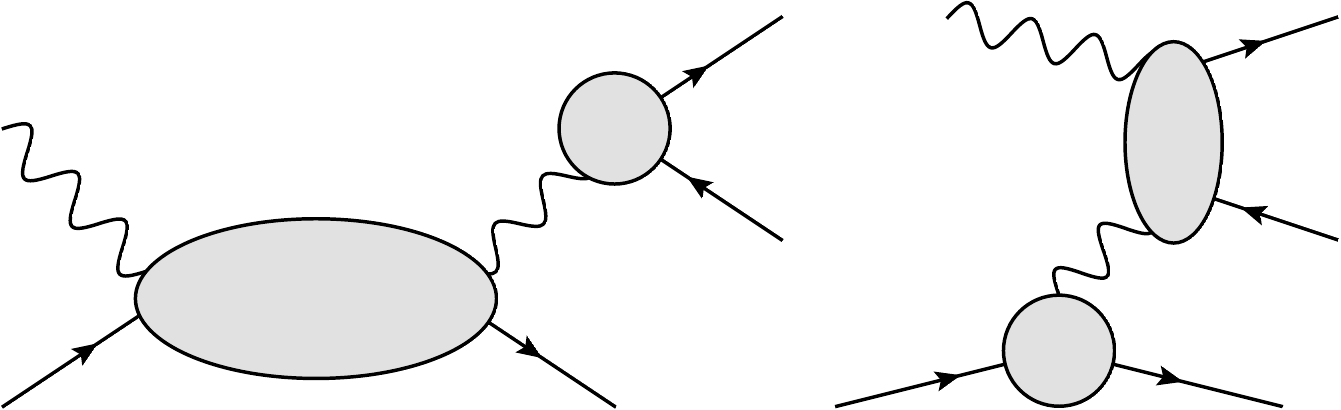}
	\caption{TCS (left panel) and BH (right panel) processes. The blob on the proton line denotes the nucleon structure, whereas the blob on the lepton line denotes the QED amplitude, including radiative corrections.}
	\label{fig:tree}
\end{figure}

In this work we consider the process 
\begin{equation}
    \gamma(q)+N(p)\rightarrow l^-(l_-)+l^+(l_+) + N(p^\prime),
    \label{eq:reaction}
\end{equation}
where the quantities in  brackets denote the four-momenta of the particles. We distinguish between two different contributions to (\ref{eq:reaction}), which are called the Bethe-Heitler (BH) process and the timelike Compton scattering process (TCS). We show the corresponding Feynman diagrams in Fig \ref{fig:tree}. 

The process (\ref{eq:reaction}) is defined in terms of three kinematic invariants:
\begin{equation}
   (p+q)^2=W^2,\qquad (l_++l_-)^2 = s_{ll},\qquad (p^\prime - p)^2=t,
\end{equation}
and two angles $\theta_l$ and $\phi_l$, which are defined in the rest-frame of the di-lepton pair, with the polar angle $\theta_l$ defined relative to the c.m. direction of $q^\prime \equiv l_+ + l_-$.

The matrix element of the BH process at leading order is given by:
\begin{align}
\mathcal{M}^{\text{BH}}_0&=\frac{ie^3}{t}\bar{u}(l_-)\biggl[\gamma^{\mu}\frac{(\sla{l_-}-\sla{q}+m)}{(l_--q)^2-m^2}\gamma^{\nu}\nonumber\\
&+\gamma^{\nu}\frac{(\sla{q}-\sla{l_+}+m)}{(q-l_+)^2-m^2}\gamma^{\mu}\biggl]v(l_+)\nonumber\\
&\times\varepsilon_{\mu}(q)\bar{N}(p^{\prime})\Gamma_{\nu}(t)N(p),
\end{align}
where $m$ denotes the mass of the lepton and where the electromagnetic vertex $\Gamma_{\nu}$ for the proton is expressed as
\begin{equation}
\Gamma_{\nu}(t)=F_D(t)\gamma_{\nu}+iF_P(t)\frac{\sigma_{\nu\alpha}\Delta^{\alpha}}{2M},\label{eq:protonVertex}
\end{equation} 
with momentum transfer to the proton $\Delta \equiv p^\prime - p$, satisfying $\Delta^2 = t$, with $M$ the proton's mass, and with 
$F_D$ ($F_P$) the Dirac (Pauli) proton form factors (FFs)  respectively.

The general TCS matrix element is given by:
\begin{align}
\mathcal{M}^{\text{TCS}}_0&=- \frac{ie^3}{s_{ll}}\bar{N}(p^{\prime})M^{\mu\nu}N(p)\varepsilon_{\mu}(q)\bar{u}(l_-)\gamma_\nu v(l_+),\label{eq:Mdvcs}
\end{align}
where $M^{\mu\nu}$ is the Compton tensor which will be specified below, and which depends on the model used to describe the interaction with the proton.

The unpolarized, fully differential cross section $d\sigma_0$ is given by
\begin{align}
\frac{d\sigma_0}{dt\,ds_{ll}\,d\Omega_{ll}^\ast}
=&\frac{1}{(2\pi)^4}\frac{1}{64}\frac{\beta}{(2M E_{\gamma})^2}
\overline{\sum_{i}}\sum_{f}\left|\mathcal{M}_0 \right|^2,\label{CrossSectionSum}
\end{align}
where $E_{\gamma}$ is the lab energy of the initial photon, which is related to $W$ as $E_\gamma = (W^2 - M^2)/(2 M)$, and $\Omega_{ll}^\ast$ is the solid angle of the lepton pair in the $l^+l^-$ center-of-mass frame, in which the lepton velocity is denoted by
\begin{equation}\label{EqBeta}
\beta=\sqrt{1-\frac{4m^2}{s_{ll}}}.
\end{equation}
The tree-level amplitude $\mathcal{M}_0$ is given by the sum of BH and TCS amplitudes:
\begin{equation}
    \mathcal{M}_0 = \mathcal{M}_0^\text{BH}+\mathcal{M}_0^\text{TCS}.
\end{equation}
In Eq. \eqref{CrossSectionSum}, we average over all polarizations in the initial state and sum over the polarizations in the final state.

\section{Models for the doubly virtual Compton amplitude}
\label{sec:compton}
\begin{figure}[h]
	\centering
  \includegraphics[scale=0.6]{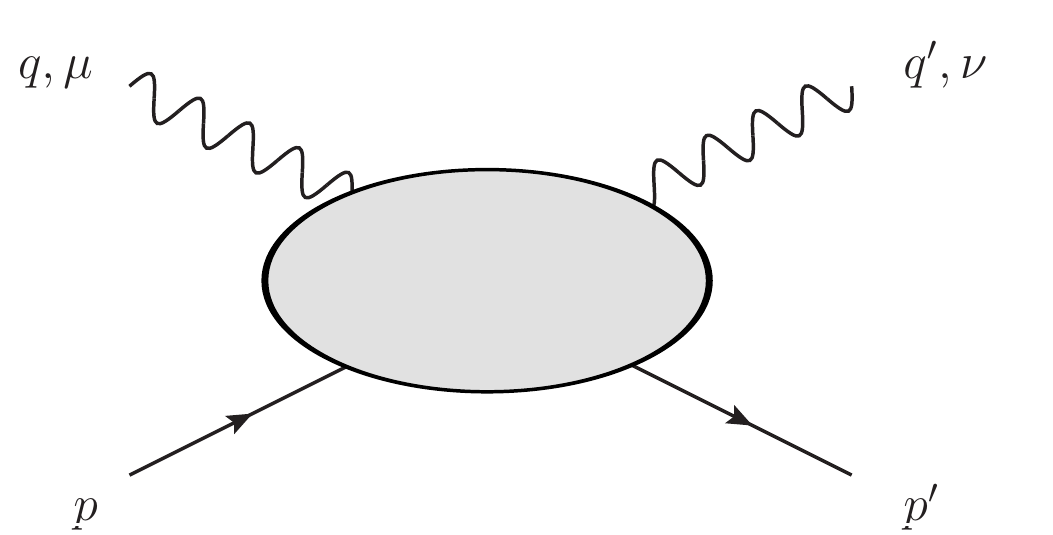}
	\caption{Diagram representing the doubly virtual Compton process. \label{fig:compton}}
\end{figure}

The doubly virtual Compton tensor $M^{\mu\nu}$ entering Eq.~(\ref{eq:Mdvcs}) is calculated from the process
\begin{equation}
    \gamma^*(q)+N(p)\rightarrow \gamma^*(q^\prime) + N(p^\prime).
\end{equation}
We show the Feynman diagram for this process in Fig. \ref{fig:compton}. The blob in this diagram represents the interaction of the incoming and outgoing photons with the nucleon. 
In the following we use the average photon ($\bar{q}$) and proton ($P$) momenta,
\begin{equation}
    \bar{q}=\frac{1}{2}(q+q^\prime),\qquad P=\frac{1}{2}(p+p^\prime).
\end{equation}

Although in this paper we study the process for a real initial photon, we will indicate below the extension to the general case of two off-shell photons.

The general doubly virtual Compton tensor $ M^{\mu\nu}$ can be constructed using $q^\mu$, $q^{\prime \mu}$, $p^\mu$, $g^{\mu\nu}$ and $\gamma^\mu$ as building blocks. From these blocks, one finds $34$ independent tensors with two indices~\cite{Tarrach:1975tu}. 
Using gauge invariance it was shown that the number of independent amplitudes can be reduced from $34$ to $18$~\cite{Tarrach:1975tu}. 
The latter number corresponds with the minimal number of helicity amplitudes for a parity conserving process, which can be determined by accounting for the possible helicity states of the photons ($3$) and fermions ($2$). 
However it was realized in Ref.~\cite{Tarrach:1975tu}, that there is in general a problem in such representation. For specific kinematical points the $18$ tensors become linearly dependent and therefore do not form a basis at these specific points anymore. As a result the corresponding Compton amplitudes display kinematic singularities at these points. To bypass this problem, Tarrach introduced an overcomplete basis by introducing three additional tensors which do not have any kinematical constraints and are valid in the whole phase space.
It was realized in Ref.~\cite{Drechsel:1997xv} that the kinematic singularities and constraints of the Compton amplitude in a minimal basis are due to the Born terms, in which the intermediate state in the Compton process in Fig.~\ref{fig:compton} is a nucleon, and that for the non-Born contributions a minimal tensor basis consisting of 18 structures  free of kinematical singularities and constraints exists.  

In order to specify the doubly virtual Compton amplitude, we need to model the internal structure of the nucleon. In this work, we will consider two different models, which are tailored for applications in two different energy regimes. In a low-energy model, which is motivated for applications to describe the hadronic structure in precision atomic physics measurements such as the Lamb shift or hyperfine splitting in muonic Hydrogen, we consider the photons to interact with the nucleon and its lowest excitation, the $\Delta(1232)$ resonance. In a high-energy model, in which at least one of the photons is highly virtual, we use perturbative QCD which allows to factorize the Compton process on the nucleon in terms of a Compton amplitude on the quark convoluted with the amplitude 
to find the quarks inside the nucleon. The latter is parameterized through GPDs.

\subsection{Born term and $\Delta$-pole model at low energies\label{sec:dvcs_low_energy}}
\begin{figure}[h]
	\centering
	\includegraphics[scale=0.6]{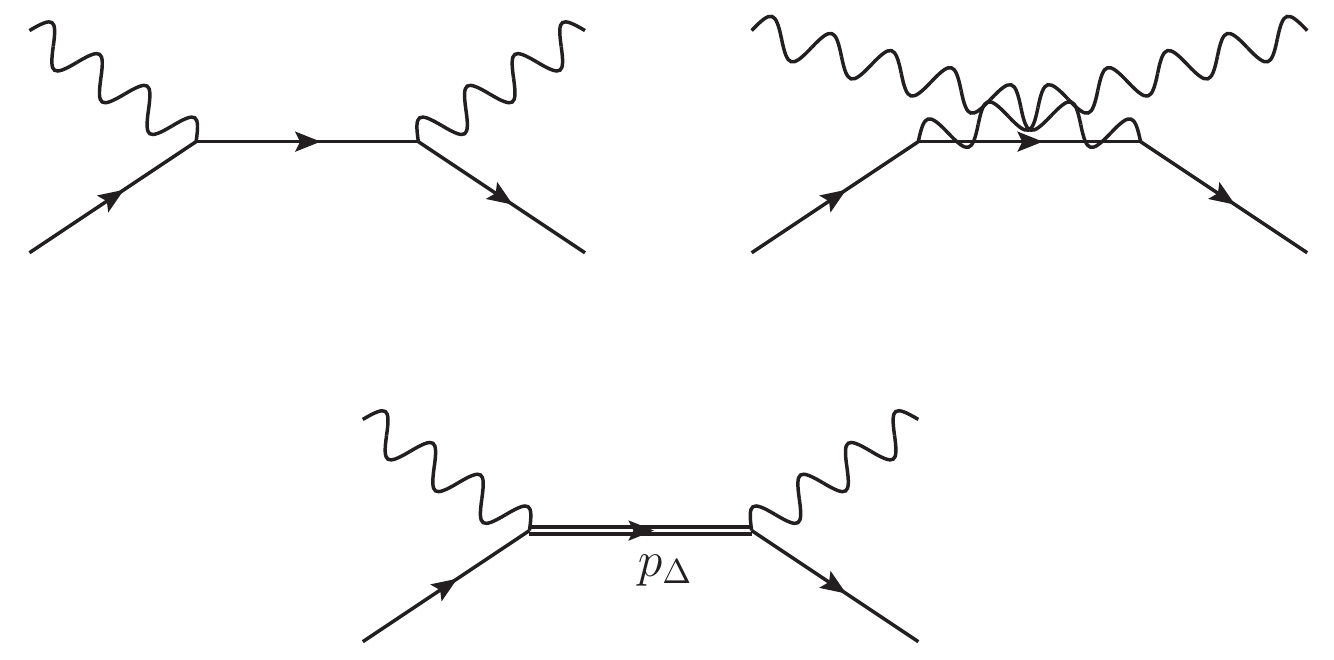}

	\caption{Born contribution (upper panel) and  $s$-channel $\Delta$-pole contribution (lower panel) to the Compton amplitude. While for the Born contribution only the sum of $s$ and $u$- channel diagrams is gauge invariant, the $s$-channel $\Delta$-pole contribution is gauge invariant by itself.}
	\label{fig:dvcs_low}
\end{figure}

At low photon energies, the leading Born (B) contribution to the Compton amplitude is described by two Feynman diagrams, 
shown in Fig.~\ref{fig:dvcs_low} (upper panel), in which a nucleon is propagating between both photon interactions. Its contribution to the tensor in Eq.~(\ref{eq:Mdvcs}) can be calculated as:
\begin{equation}
    M_{\rm B}^{\mu\nu}=\Gamma_f^\nu\frac{\sla{p}+\sla{q}+M}{(p+q)^2-M^2}\Gamma_i^\mu+\Gamma_i^\mu\frac{\sla{p}'-\sla{q}+M}{(p'-q)^2-M^2}\Gamma_f^\nu,
\end{equation}
where $\Gamma^\mu_i$ ($\Gamma^\nu_f$) are the initial (final) state proton vertices, given by analogous expressions as  Eq.~\eqref{eq:protonVertex} in which $\Delta$ is replaced by $q$ ($-q^\prime$) for $\Gamma^\mu_i$ ($\Gamma^\nu_f$) respectively. Note that the FFs entering $\Gamma^\nu_f$ correspond with a timelike virtuality. 
For the numerical evaluation of these FFs we use the paramaterization of Ref.~\cite{Lomon:2012pn}. This parameterization allows the analytical continuation based on dispersion relations into the unphysical part of the timelike region, $0 < q^{\prime 2} < 4 M^2$, in which no direct experimental extraction exists.

In addition to the Born term, we also evaluate the matrix element of the leading contribution due to the $\Delta$ resonance in the general case with two off-shell photons. The corresponding $s$-channel diagram is shown in Fig.~\ref{fig:dvcs_low} (lower panel) and its contribution to the tensor in Eq.~(\ref{eq:Mdvcs}) can be calculated as:
\begin{align}
  M_{s\Delta}^{\mu\nu} =&  \tilde{\Gamma}_{\gamma N \Delta}^{\alpha\nu}(p^\prime,p+q)\frac{(\sla{p}+\sla{q}+M_\Delta)(-g_{\alpha\beta}+\frac{1}{3}\gamma_\alpha\gamma_\beta)}{W^2 - M_\Delta^2+i M_\Delta\Gamma_\Delta(W^2)}\nonumber\\
&\times\Gamma_{\gamma N \Delta}^{\beta\mu}(p+q,p).\label{mdelta}
\end{align}
\begin{figure}[h]
	\centering
  \includegraphics[scale=0.6]{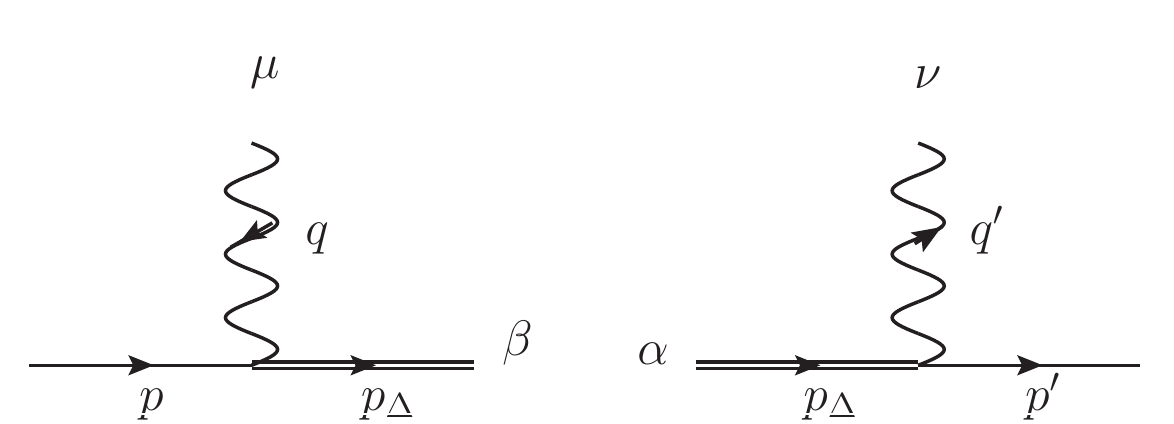}
	\caption{The $\gamma^* N \Delta$ vertex $\Gamma_{\gamma N \Delta}^{\beta\mu}$ (left diagram) and its adjoint, $\tilde{\Gamma}_{\gamma N \Delta}^{\alpha\nu}$ (right diagram). }
	\label{d_vertex}
\end{figure}

In Eq. \eqref{mdelta}, $\Gamma_{\gamma N \Delta}^{\beta\mu}$ and  $\tilde{\Gamma}_{\gamma N \Delta}^{\alpha\nu}$ refer to the $\gamma^* N \Delta$ vertex function and its adjoint respectively. They are shown in Fig. \ref{d_vertex} and can be expressed in terms of three transition form factors as~\cite{Pascalutsa:2006up}:
\begin{eqnarray}
\Gamma^{\beta\mu}_{\gamma N \Delta}(p_\Delta,p)
&=& \sqrt{\frac{3}{2}}\frac{(M_\Delta+M)}{M Q_+^2}\biggl\{g_M(q^2)i\epsilon^{\beta\mu\kappa\lambda}(p_\Delta)_\kappa q_\lambda\nonumber\\
&&- g_E(q^2)(q^\beta p_\Delta^\mu-q\cdot p_\Delta g^{\beta \mu})\gamma_5\nonumber\\
&&- g_C(q^2) \frac{1}{M_\Delta} \left[\sla{p}_\Delta(q^\beta q^\mu-q^2g^{\beta\mu}) \right. \nonumber \\
&&\left. \hspace{1cm}-\gamma^\beta(q\cdot p_\Delta q^\mu-q^2p^\mu_\Delta) \right]\gamma_5\biggl\},
\label{eq:ndelspace}
\end{eqnarray}
and its adjoint:
\begin{eqnarray}
\tilde{\Gamma}^{\alpha\nu}_{\gamma N \Delta}(p^\prime,p_\Delta) &=& -\sqrt{\frac{3}{2}}\frac{(M_\Delta+M)}{M Q_+^{\prime 2}}\biggl\{g_M(q^{\prime 2}) i\epsilon^{\alpha\nu\kappa\lambda}(p_\Delta)_\kappa q^\prime_\lambda\nonumber\\
&&-g_E(q^{\prime 2})(q^{\prime \alpha} p_\Delta^\nu-q^\prime \cdot p_\Delta g^{\alpha\nu})\gamma_5
\nonumber \\
&&-g_C(q^{\prime 2}) \frac{1}{M_\Delta} \gamma_5 \left[\sla{p}_\Delta(q^{\prime \alpha} q^{\prime \nu}-q^{\prime 2} g^{\alpha\nu}) \right. \nonumber \\
&&\hspace{1cm}\left. -\gamma^\alpha(q^\prime \cdot p_\Delta q^{\prime \nu} - q^{\prime 2} p^\nu_\Delta) \right]\biggl\},
\label{eq:ndeltime}
\end{eqnarray}
where we defined $Q_\pm=\sqrt{(M_\Delta\pm M)^2 - q^2}$ and likewise 
$Q^\prime_\pm=\sqrt{(M_\Delta\pm M)^2 - q^{\prime 2}}$. Note that the FFs $g_M$, $g_E$, and $g_C$ appearing in Eq.~(\ref{eq:ndelspace}) have spacelike virtuality ($q^2 < 0$), whereas the corresponding ones in the adjoint vertex of Eq.~(\ref{eq:ndeltime}) have timelike virtuality ($q^{\prime 2} > 0$). 

The form factors $g_M$, $g_E$, and $g_C$ can be expressed by the more conventional magnetic dipole ($G_M^*$), electric quadrupole ($G_E^*$), and Coulomb quadrupole ($G_C^*$) transition FFs as:
\begin{align}
    g_M&=\frac{Q_+}{M+M_\Delta}(G_M^*-G_E^*),\nonumber\\
    g_E&=-\frac{Q_+}{M+M_\Delta}\frac{2}{Q_-^2}\{(M_\Delta^2-M^2+q^2)G_E^*-q^2G_C^*\},\nonumber\\
    g_C&=\frac{Q_+}{M+M_\Delta}\frac{1}{Q_-^2}\{4M_\Delta^2G_E^*-(M_\Delta^2-M^2+q^2)G_C^*\},
\end{align}
with the so-called Ash FFs parameterized, for spacelike virtuality $Q^2 = -q^2$, through the MAID2007 analysis as~\cite{Drechsel:2007if,Tiator:2011pw}:
\begin{align}
    G^{*}_{M}(Q^2)&=3.00 (1+0.01 Q^2)e^{-0.23 Q^2}G_D(Q^2),\nonumber\\
    G^{*}_{E}(Q^2)&=0.064 (1-0.021 Q^2)e^{-0.16 Q^2}G_D(Q^2),\nonumber\\
    G^{*}_{C}(Q^2)&=0.124 \frac{1+0.12 Q^2}{1+4.9Q^2/(4M^2)}\frac{4M_\Delta^2e^{-0.23 Q^2}G_D(Q^2)}{M_\Delta^2-M^2} ,
\end{align}
with $Q$ in GeV and the dipole FF $G_D(Q^2)=1/(1+Q^2/0.71)^2$. For small timelike virtualities, $0 < q^{\prime 2} < (M_\Delta - M)^2$, we can extrapolate the expressions for spacelike virtualities by the substitution $Q^2 \to - q^{\prime 2}$.

\subsection{High-energy timelike Compton Scattering in terms of GPDs\label{sec:dvcs_high_energy}}

\begin{figure}[h]
\centering
  \includegraphics[scale=0.55]{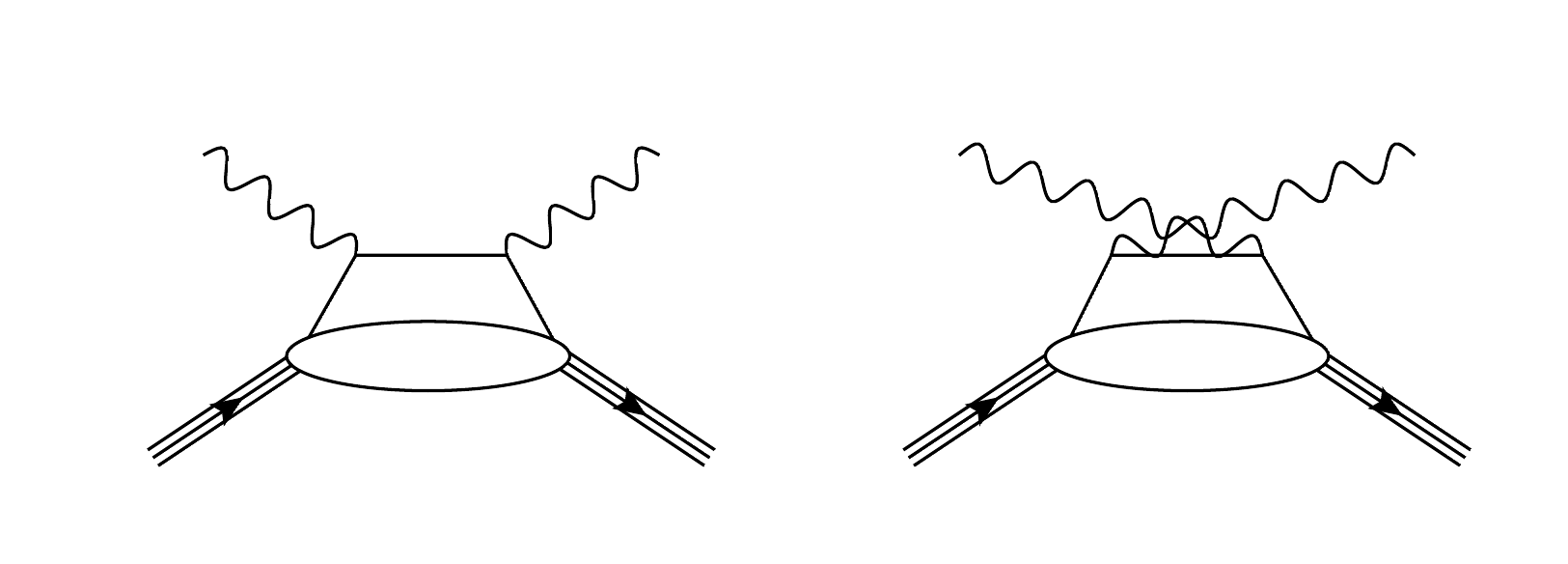}
  \caption{Handbag diagrams for the doubly virtual Compton amplitude. The 
  single (composite) lines represent quarks (nucleons) respectively. The blobs represent the GPDs. 
  \label{dia:handbag}}
\end{figure}

For deeply-virtual Compton scattering, in which at least one of the photons has a large virtuality, we can express the doubly virtual Compton tensor using perturbative QCD in terms of GPDs. 
At leading order in the large virtuality, the deeply virtual Compton tensor can be calculated through the handbag diagrams, shown in Fig.~\ref{dia:handbag}. These handbag diagrams express the factorization of the process in terms of the Compton amplitude on the quark convoluted with the amplitude to find the quark in the nucleon, which is parameterized through the GPDs.

To describe the deeply-virtual Compton process with two virtual photons, it is convenient to define the Lorentz invariants $\xi$ and $\xi'$ as:
\begin{eqnarray}
    \xi&\equiv&-\frac{\Delta\cdot \bar{q}}{2P\cdot \bar{q}} 
    = \frac{-q^2 + q'^2}{2 (W^2 - M^2) -  q^2 - q'^2 + t} , 
    \label{eq:xi}\\
    \xi^\prime &\equiv& -\frac{\bar{q}^2}{2P\cdot \bar{q}} 
    = \frac{-q^2 - q'^2 + t/2}{2 (W^2 - M^2) -  q^2 - q'^2 + t}.
    \label{eq:xip}
\end{eqnarray}

To calculate the handbag diagrams, we first express the four-momenta $P^\mu$ and $\bar q^\mu$ in terms of the lightlike four-vectors $\tilde{p}$ and $n$, with $\tilde p \cdot n = 1$, as:
\begin{eqnarray}
P^\mu &=& \tilde p^\mu + \frac{\bar M^2}{2} n^\mu, \label{eq:P}\\
\bar q^\mu &=& - \tilde \xi^\prime \tilde p^\mu - \frac{\bar q^2}{2 \tilde \xi^\prime} n^\mu \label{eq:qbar}, 
\end{eqnarray}
with $\bar M^2 = M^2 - t / 4$. 
The variables $\tilde{\xi}$ and $\tilde{\xi}'$ are related to the invariants $\xi$ and $\xi^\prime$ introduced in Eqs.~(\ref{eq:xi},\ref{eq:xip}) as:
\begin{align}
\tilde{\xi}&=\xi\frac{1+\tilde{\xi}'^2 \bar{M}^2/ \bar{q}^2}{1-\tilde{\xi}'^2\bar{M}^2/\bar{q}^2}, 
\label{eq:tildexi} \\
    \tilde{\xi}'&=\xi'\frac{2}{1+\sqrt{1-4\xi'^2\bar{M}^2/ \bar{q}^2}}.
    \label{eq:tildexip}
\end{align}
We notice that the difference between the tilded quantities 
$\tilde{\xi}, \tilde{\xi^\prime}$ of Eqs.~(\ref{eq:tildexi}, \ref{eq:tildexip}) and the quantities $\xi, \xi^\prime$ of 
Eqs.~(\ref{eq:xi}, \ref{eq:xip}) involve kinematical corrections due to the target mass $M$ and momentum tranfer $t$.
In the following, we will consider the Bjorken limit $\bar{q}^2\gg \bar{M}^2$, in which:
\begin{equation}
    \tilde{\xi} \rightarrow \xi,\qquad \tilde{\xi}' \rightarrow \xi'.
\end{equation}

With these kinematic definitions, the double deeply virtual Compton scattering (DDVCS) tensor at leading twist-2 
can be expressed as~\cite{Guidal:2002kt}:
\begin{eqnarray}
M^{\mu\nu}_{\rm{DDVCS}} &=& \frac{1}{2}(-g_{\mu\nu})_{\perp}\int_{-1}^{1}dx
\left[\frac{1}{x-\xi^\prime+i\epsilon}+\frac{1}{x+\xi^\prime-i\epsilon}\right]\nonumber\\
&\times& \left\{H(x,\xi,t)\sla{n}+E(x,\xi,t)i\sigma^{\alpha\beta}n_\alpha \frac{\Delta_\beta}{2M}\right\}\nonumber\\
&+&\frac{i}{2}(\varepsilon_{\nu\mu})_{\perp}\int_{-1}^{1}dx
\left[\frac{1}{x-\xi^\prime+i\epsilon}-\frac{1}{x+\xi^\prime-i\epsilon}\right]\nonumber\\
&\times& \left\{\tilde{H}(x,\xi,t)\sla{n}\gamma_5+\tilde{E}(x,\xi,t)\gamma_5\frac{\Delta\cdot n}{2M}\right\},
\label{eq:DDVCS}
\end{eqnarray}
where
\begin{align}
    (-g_{\mu\nu})_{\perp}&=-g_{\mu\nu}+\tilde{p}_\mu n_\nu+\tilde{p}_\nu n_\mu,\nonumber\\
    (\varepsilon_{\nu\mu})_\perp&=\varepsilon_{\nu\mu\alpha\beta}n^\alpha \tilde{p}^\beta,
\end{align}
and where the lightlike four-vectors $\tilde p$ and $n$ are obtained from Eqs.~(\ref{eq:P},\ref{eq:qbar}) as:
\begin{align}
    n^\mu&=\frac{1}{\tilde{\xi}^\prime \bar{M}^2/2-\bar{q}^2/(2\tilde{\xi}^\prime)}\left\{\tilde{\xi}^\prime P^\mu +\bar{q}^\mu\right\},\nonumber\\
    \tilde{p}^\mu&=\frac{-1}{\tilde{\xi}^\prime \bar{M}^2-\bar{q}^2/\tilde{\xi}^\prime}\left\{\bar{q}^2/\tilde{\xi}^\prime P^\mu +\bar{M}^2\bar{q}^\mu\right\}.
\end{align}
Furthermore in Eq.~(\ref{eq:DDVCS}), $H$, $E$, $\tilde H$, and $\tilde E$ are the GPDs, which depend on the two quark momentum fractions $x$ and $\xi$, and on the momentum transfer $t$. 

One can apply the above formula of Eq.~(\ref{eq:DDVCS}) for the DDVCS tensor to two experimentally important limits. 
The first is the deeply-virtual Compton scattering (DVCS) process, in which the final photon is real ($q^{\prime 2}$ = 0), and the intial photon's virtuality is large, i.e. $Q^2 = -q^2 \gg -t$, for which one has: 
\begin{eqnarray}
{\rm DVCS:} \quad \quad 
\xi = \xi^\prime = \frac{x_B/2}{1 - x_B/2}, 
\end{eqnarray}
with Bjorken variable 
$x_B \equiv Q^2/(2 p \cdot q)$. 
The second limit is the timelike Compton scattering (TCS) process 
with inital photon real 
($q^2$ = 0), and large timelike virtuality, i.e. $q^{\prime 2} \gg -t$, for which one has:
\begin{eqnarray}
{\rm TCS:} \quad \quad 
\xi = -\xi^\prime = \frac{q^{\prime 2}}{4 M E_\gamma - q^{\prime 2}}.
\end{eqnarray}
The TCS amplitude can then be obtained by using the expression for the 
DDVCS tensor of Eq.~(\ref{eq:DDVCS}) in the TCS limit in Eq.~(\ref{eq:Mdvcs}). 

In the following, we will consider the observables for the TCS process on an unpolarized nucleon at small momentum transfer $-t \ll q^{\prime 2}$. For these observables, the dominant contribution arises from the structure function $H$, which is normalized to the proton Dirac FF $F_D(t)$. When studying the influence of the radiative corrections on the TCS observables at small values of $-t$, we will therefore neglect the contribution of the GPDs $E$, $\tilde{H}$ and $\tilde{E}$ in our study below. 
For the numerical evaluation in this work, we will use the GPD parametrizations from the
VGG model~\cite{Vanderhaeghen:1998uc, Vanderhaeghen:1999xj,Goeke:2001tz,Guidal:2004nd}, summarized in Ref.~\cite{Guidal:2013rya} as:
\begin{equation}
    H(x,\xi,t)=\frac{4}{9} H^u_{\rm{DD}}(x,\xi,t)+\frac{1}{9}H^d_{\rm{DD}}(x,\xi,t)+D(\frac{x}{\xi},t).
    \label{eq:gpdmodel}
\end{equation}
The parameterization is based on a double-distribution (DD) ansatz for the ($x$,$\xi$)-dependence of the up (down) quark~\footnote{The small $s$-quark GPD contribution is neglected in this work.} GPDs $H^u_{\rm{DD}}$ ($H^d_{\rm{DD}}$), 
with parameter values $b_v = 1$ ($b_s = 5$) for valence (sea) quarks respectively,
and on a Reggeized ansatz for the $t$-distribution, which was found to give a global description of existing DVCS data~\cite{Dupre:2016mai,Dupre:2017hfs}. Furthermore, we added an isoscalar  so-called $D$-term contribution in Eq.~(\ref{eq:gpdmodel}), which only depends on the two variables $x/\xi$ and $t$, and which is directly related to the subtraction function in a dispersive framework for the Compton amplitude. For its evaluation, we use the dispersive estimate of Ref.~\cite{Pasquini:2014vua}.

\section{Vacuum polarization at first order\label{sec:vac-pol}}
\begin{figure}
	\includegraphics[scale=0.6]{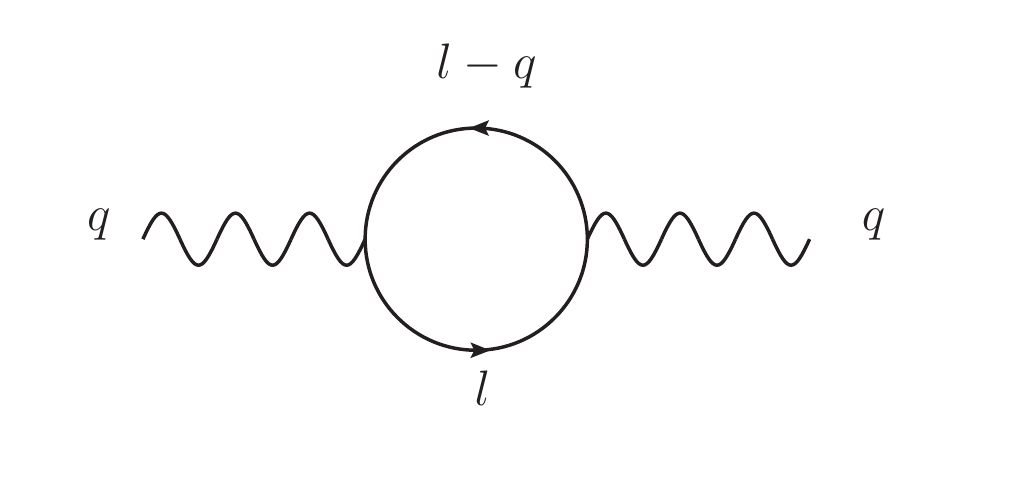}\hspace{0.cm}
	\caption{Vacuum polarization diagram. The fermion loop can be either electrons or muons.\label{Vacuumdiagram}}
\end{figure}

We start our discussion of the first order radiative corrections with vacuum polarization process which is shown to first order in Fig. \ref{Vacuumdiagram}. The photon propagator can be written as:
\begin{equation}
    D^{\mu\nu}(q)=D_0^{\mu\nu}(q)+D_0^{\mu\alpha}(q)\Pi_{\alpha\beta}(q)D_0^{\beta\nu}(q),
\end{equation}
where $D_0^{\mu\nu}$ is the leading order photon propagator
\begin{equation}
    D_0^{\mu\nu}=\frac{-g^{\mu\nu}}{q^2},
\end{equation}
and $\Pi_{\alpha\beta}$ is the vacuum polarization, which is given at first order in $\alpha \equiv e^2 / (4 \pi)$ by:
\begin{equation}
    \Pi^{\mu\nu}(q)=-ie^2\int\frac{d^4l}{(2\pi)^4}\frac{\text{Tr}\left[\gamma^{\mu}(\sla{l}-\sla{k}+m_l)\gamma^{\nu}(\sla{l}+m_l)\right]}{[(l-k)^2-m_l^2][l^2-m_l^2]}
\end{equation}
where $m_l$ is the mass of the lepton in the loop. Due to gauge-invariance, $q_{\mu}\Pi^{\mu\nu}=q_{\nu}\Pi^{\mu\nu}=0$, and the vacuum polarization can be decomposed as:
\begin{equation}
    \Pi^{\mu\nu}(q)=(-g^{\mu\nu}q^2+q^{\mu}q^{\nu})\Pi(q^2).
\end{equation}
The scalar function $\Pi(q^2)$ has an UV divergence. In dimensional regularisation it is given by~\cite{Vanderhaeghen:2000ws}:
\begin{align}
    \Pi(q^2)=&-\frac{\alpha}{3\pi}\biggl[\frac{1}{\epsilon_{\text{UV}}}-\gamma_E+\ln\left(\frac{4\pi\mu^2}{m_l^2}\right)-\left(v^2-\frac{8}{3}\right)\nonumber\\&+\frac{v}{2}(v^2-3)\ln\left(\frac{v+1}{v-1}\right)\biggl],\label{vacuumunren}
\end{align}
where we defined $v^2\equiv1- 4m_l^2 / q^2$. In the following we consider muons and electrons in the fermion loop.

The UV-divergence (in limit $\epsilon_{\text{UV}} \to 0+$) in Eq. \eqref{vacuumunren} is removed by the renormalization constant $Z_3$:
\begin{equation}
    \tilde{\Pi}(q^2)=\Pi(q^2)-(Z_3-1).
\end{equation}
In the on-shell scheme this constant is fixed by requiring, that the renormalized vacuum polarization $\tilde{\Pi}(q^2)$ has a pole with residue $1$ at $q^2=0$:
\begin{equation}
    Z_3=1+\Pi(q^2=0).
\end{equation}
The renormalized vacuum polarization is then given by:
\begin{equation}
    \tilde{\Pi}(q^2)=\frac{\alpha}{3\pi}\left[\left(v^2-\frac{8}{3}\right)+\frac{v}{2}(3-v^2)\ln\left(\frac{v+1}{v-1}\right)\right],\label{EqVPRen}
\end{equation}
and the renormalized photon propagator by:
\begin{equation}
    \tilde{D}^{\mu\nu}(q)=\frac{-g^{\mu\nu}}{q^2}\left[1+\tilde{\Pi}(q^2)\right]+\frac{q^{\mu}q^{\nu} \tilde{\Pi}(q^2)}{q^4}.
\end{equation}
Note that, due to gauge invariance, only the term proportional to $g^{\mu\nu}$ contribute to the BH and TCS amplitudes, such that the corrections factorize as:
\begin{align}
\mathcal{M}^{\text{TCS}}_\text{vac pol}&=\tilde{\Pi}(s_{ll})\mathcal{M}^{\text{TCS}},\nonumber\\
\mathcal{M}^{\text{BH}}_\text{vac pol}&=\tilde{\Pi}(t)\mathcal{M}^{\text{BH}}.
\label{eq:tcsvac}
\end{align}
For the evaluation of $\tilde{\Pi}(s_{ll})$ we need to perform an analytic continuation of Eq.~(\ref{EqVPRen}):
\begin{align}
&\tilde{\Pi}(s_{ll})= \frac{\alpha}{3\pi}\left(v^2-\frac{8}{3}\right)\nonumber\\&+\frac{\alpha}{3\pi}
\begin{cases}    \frac{\tilde{v}}{2}(v^2-3)\left[2\arctan{\tilde{v}}-\pi\right]   & 0<s_{ll}<4m_l\\
  \frac{v}{2}(3-v^2)\left[\ln\left(\frac{1+v}{1-v}\right)-i\pi\right]& s_{ll}\geq 4 m_l^2,
\end{cases}
\end{align}
where $\tilde{v}\equiv iv=\sqrt{4m_l^2/s_{ll}-1}$.

\section{Vertex correction to the TCS amplitude}
\label{sec:one_loop_dvcs}

\begin{figure}[h]
	\centering
  \includegraphics[scale=0.6]{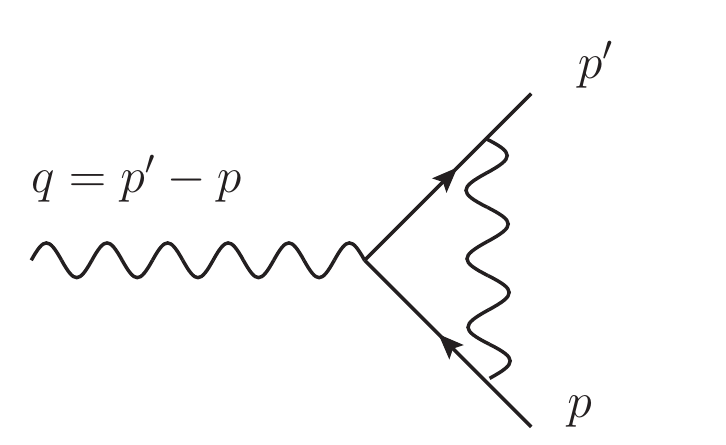}
	\caption{One-loop QED correction to the vertex.}
	\label{3pointgraphs}
\end{figure}

We next consider the one-loop QED corrections to the process
\begin{equation}
    \gamma^*(q)\rightarrow l(p)+\bar{l}(p^\prime),
\end{equation}
shown in Fig.~\ref{3pointgraphs}.

The corresponding matrix element can be expressed in terms of two form factors, $F^e_D$ and $F^e_P$ called Dirac and Pauli electron form factors respectively:
\begin{equation}
    \Gamma^\mu(q^2)= F^e_D(q^2) \gamma^\mu+i F^e_P(q^2) \sigma^{\mu\nu}\frac{q_\nu}{2m}.
\end{equation}
To first order in $\alpha$, the Dirac form factor $F^e_D$ is divergent, such that a regularisation procedure is needed. In dimensional regularisation it can be expressed as~\cite{Vanderhaeghen:2000ws}:
\begin{align}
&F^e_D(q^2)\nonumber\\
&=\left(\frac{\alpha}{4\pi}\right)\left\{\left[\frac{1}{\epsilon_\text{UV}}-\gamma_E+\ln\left(\frac{4\pi\mu^2}{m^2}\right)\right]\right.\nonumber\\
&\left.+\left[\frac{1}{\epsilon_\text{IR}}-\gamma_E+\ln\left(\frac{4\pi\mu^2}{m^2}\right)\right]\frac{1+v^2}{v}\ln\left(\frac{v+1}{v-1}\right)\right.\nonumber\\
&\left.+\frac{2v^2+1}{v}\ln\left(\frac{v+1}{v-1}\right)+\frac{v^2+1}{2v}\ln\left(\frac{v+1}{v-1}\right)\ln\left(\frac{v^2-1}{4v^2}\right)\right.\nonumber\\
&\left.+\frac{1+v^2}{v}\left[\text{Li}_2\left(\frac{v+1}{2v}\right)-\text{Li}_2\left(\frac{v-1}{2v}\right)\right]\right\}.
\label{eq:FDe}
\end{align}
The Pauli form factor $F^e_P$ is finite and is given by:
\begin{align}
F^e_P(q^2)&=-\frac{\alpha}{4\pi}\frac{v^2-1}{v}\ln\left(\frac{v-1}{v+1}\right),\label{eq:pauli}
\end{align}
where $v$ is defined in the same way as below Eq. \eqref{vacuumunren} by replacing the mass $m_l\rightarrow m$.

As can be seen from Eq.~\eqref{eq:FDe}, $F^e_D$ has an ultraviolet (UV) divergence (in limit $\epsilon_{\text{UV}} \to 0+$), as well as an infrared divergence (in limit $\epsilon_{\text{IR}} \to 0-$). The UV divergence gets removed by the on-shell subtraction scheme, in which the vertex counter term is defined to fix the electron charge $e$ at $q^2=0$.
One finds at $q^2=0$ the renormalization constant:
\begin{align}
Z_1&=1-F^e_D(0)=\nonumber\\
&=1-\frac{\alpha}{4\pi}\biggl\{\left[\frac{1}{\epsilon_\text{UV}}-\gamma_E+\ln\left(\frac{4\pi\mu^2}{m^2}\right)\right]\nonumber\\
&+2\left[\frac{1}{\epsilon_\text{IR}}-\gamma_E+\ln\left(\frac{4\pi\mu^2}{m^2}\right)\right]+4\biggl\}. \label{onshell_vertex_counterterm}
\end{align}
This leads to the renormalized (on-shell) form factor:
\begin{equation}
\tilde{F}^e_D(q^2)=F^e_D(q^2)-F^e_D(0). \label{renormalized_QED_F1}
\end{equation}
For the case when the momentum transfer $q^2$ becomes timelike, i.e. $q^2>0$, one has to perform an analytic continuation of both form factors.

To calculate the corrections to the dVCS matrix element, we just have to contract $\Gamma^\mu$ with the dVCS tensor. Adding the vacuum polarization correction of Eq.~(\ref{eq:tcsvac}), this yields the one-loop radiative correction to the TCS amplitude as:
\begin{align}
\mathcal{M}^{\text{TCS}}_\text{1-loop}=-&\frac{ie^3}{s_{ll}}\bar{N}(p^{\prime})M^{\mu\nu} (t)N(p)\varepsilon_{\mu}(q)\bar{u}(l_-)\tilde{\Gamma}_\nu(s_{ll}) v(l_+)\nonumber\\
&+\mathcal{M}^{\text{TCS}}_\text{vac pol},
\end{align}
where $\tilde{\Gamma}_\nu$ denotes the renormalized vertex, and where we indicated that the momentum transfer which appears in the Compton tensor is given by $t$.
The remaining IR divergence in Eq.~(\ref{eq:FDe}) will be discussed in Section~\ref{sec:soft_real}. 

\section{One-loop corrections to the BH process\label{sec:BH1L}}
\label{sec:one_loop_bh}
\begin{figure}
	\centering
  \includegraphics[scale=0.47]{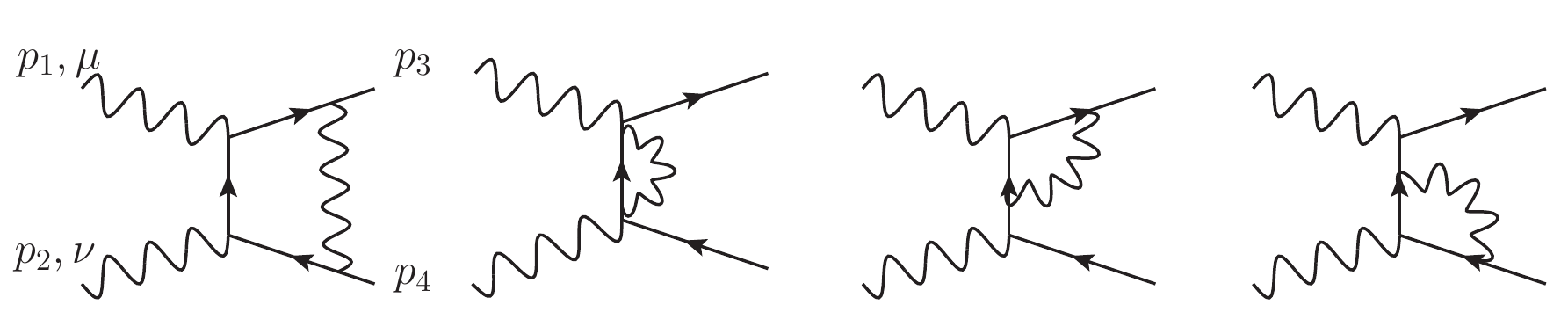}
	\caption{Diagrams contributing to the one loop corrections for the process $\gamma^*(p_1) \gamma^*(p_2) \rightarrow l(p_3) \bar{l}(p_4)$}
	\label{4pointgraphs}
\end{figure}

In order to calculate the one-loop diagrams contributing to the BH process, we consider corrections to the process: 
\begin{equation}
    \gamma^*(p_1)+\gamma^*(p_2) \rightarrow l(p_3)+\bar{l}(p_4).
\end{equation}
We show the contributing diagrams in Fig. \ref{4pointgraphs}. We use the standard definition of Mandelstam variables:
\begin{equation}
    s_{ll}=(p_1+p_2)^2,\quad t_{ll}=(p_1-p_3)^2.
\end{equation}

Using $p_1^\mu$, $p_2^\mu$, $p_3^\mu$, $g^{\mu\nu}$ and $\gamma^\mu$ as building blocks we find $34$ independent tensors with two indices to expand the one-loop amplitude. As discussed in Sec. \ref{sec:dvcs_low_energy}, we can reduce the number of amplitudes from $34$ to $21$ amplitudes, which are gauge invariant and free of any kinematic singularities, such that we can write:
\begin{equation}
    M^{\mu\nu}_{\gamma^* \gamma^* \rightarrow l \bar{l}}(p_1^2,p_2^2,s_{ll},t_{ll})=\sum_{i=1,..,21} B_i(p_1^2,p_2^2,s_{ll},t_{ll}) \tau_i^{\mu \nu},
    \label{eq:ggll_oneloop}
\end{equation}
with $\tau_i^{\mu \nu}$ the tensor basis introduced in \cite{Tarrach:1975tu}, and $B_i$ the corresponding invariant amplitudes. 

We calculated all $21$ contributing amplitudes up to the one-loop order using the same setup as in Ref. \cite{Heller:2019dyv} in dimensional regularisation. We use \texttt{QGRAF} \cite{Nogueira:1991ex} to generate all contributing diagrams, \texttt{FORM} \cite{Kuipers:2012rf} to implement the Dirac algebra and \texttt{Reduze\;2} \cite{vonManteuffel:2012np} to generate IBP identities. The amplitudes are then expressed as a sum over scalar integrals with coefficients which are rational functions in all external scales and the space-time dimension $d=4-2\epsilon$. In total we need $7$ master integrals, which are all listed in Ref. \cite{Heller:2019dyv}.

To subtract UV divergences we use the on-shell renormalization scheme. In addition to the vertex counter terms defined in Sec. \ref{sec:one_loop_dvcs}, we also need counter terms from the fermion self-energies $\Sigma(p)$, where $p$ denotes the incoming four momentum. The renormalization condition fixes the pole of $\Sigma(p)$ at $p^2=m^2$ with residue equal to one.

Besides the UV divergences, one also encounters IR divergences in the amplitude. We checked that the matrix element has the correct infrared structure after performing the UV renormalization. The infrared structure of the amplitude can be calculated using the soft-photon approximation, in which the loop momenta $l$ of the photon is assumed to scale soft, i.e. $l\sim \lambda$, where $\lambda$ is small compared to all external scales. In this approximation the calculation is performed at leading order in $\lambda$. The contribution factorizes in terms of the born amplitude: 
\begin{align}
    \mathcal{M}^{\text{BH}}_\text{soft}=&-\frac{\alpha}{2\pi}\left\{\left(s_{ll}-2m^2\right) C_0\left(m^2, s_{ll}, m^2,0, m^2, m^2\right)\right.\nonumber\\
    &\left.+\left[\frac{1}{\epsilon_\text{IR}}-\gamma_E+\ln\left(\frac{4\pi\mu^2}{m^2}\right)\right] \right\}\mathcal{M}^{\text{BH}}_0,
\end{align}
with the three-point function:
\begin{align}
C_0&\left(m^2, s_{ll}, m^2,0, m^2, m^2\right)=\frac{1}{s_{ll}\beta}\left\{\left[\frac{1}{\epsilon_\text{IR}}-\gamma_E\right.\right.\nonumber\\
&\left.\left.+\ln\left(\frac{4\pi\mu^2}{m^2}\right)\right]\ln \left(\frac{\beta-1}{\beta+1}\right)+2\;\text{Li}_2\left(\frac{\beta-1}{2\beta}\right)\right.\nonumber\\
&\left.+\ln^2\left(\frac{\beta-1}{2\beta}\right)-\frac{1}{2}\ln^2\left(\frac{\beta-1}{\beta+1}\right)-\frac{\pi^2}{6}\right\}.\label{CFunct}
\end{align}
Note that the same formula applies for the TCS matrix element in the soft-photon limit. Therefore, in this approximation the correction also factorizes on the level of the cross section. It is given by:
\begin{equation}
d\sigma_{\text{s;v}}=d\sigma_{0}\left(\vphantom{\frac{1}{2}}\delta^\text{IR}_\text{s;v}+\delta_\text{s;v}\right),
\end{equation}
with the infrared-divergent part:
\begin{align}
\delta^\text{IR}_\text{s;v}=&\left(\frac{-\alpha}{\pi}\right)\left[\left(\frac{1+\beta^2}{2\beta}\right)\ln\left(\frac{1-\beta}{1+\beta}\right)+1\right]\nonumber\\
&\times\left[\frac{1}{\epsilon_\text{IR}}-\gamma_E+\ln\left(\frac{4\pi\mu^2}{m^2}\right)\right],\label{eq:soft_div_virt}
\end{align}
and the finite part:
\begin{align}
\delta_\text{s;v}=&\left(\frac{-\alpha}{\pi}\right)\left(\frac{1+\beta^2}{2\beta}\right)
\left\{2\;\text{Li}_2\left(\frac{2\beta}{\beta+1}\right)\right.\nonumber\\
&\left.+\frac{1}{2}\ln^2\left(\frac{1-\beta}{1+\beta}\right)-\pi^2\right\}.\label{eq:soft_finite_virt}
\end{align}

As an additional check we were also able to reproduce the results of Ref. \cite{Heller:2019dyv} numerically, where the corrections to the unpolarized BH cross section have been calculated.

Using Eq. \eqref{eq:ggll_oneloop} and adding the contribution from vacuum polarisation, the one-loop matrix element corresponding to the BH process with the incoming photon on-shell can be calculated by contracting with the photon polarization vector and the proton line:
\begin{align}
\mathcal{M}^{\text{BH}}_\text{1-loop}=&\frac{ie^3}{t}\bar{u}(l_-)M^{\mu\nu}_{\gamma^* \gamma^* \rightarrow l \bar{l}}(0,t,s_{ll},t_{ll})\;
v(l_+)\varepsilon_{\mu}(q)\nonumber\\
&\times\bar{N}(p^{\prime})\Gamma_{\nu}(t)N(p)+\mathcal{M}^{\text{BH}}_\text{vac pol},
\end{align}
where we now can identify $t_{ll}=(q-l_-)^2$.

\section{Soft-photon bremsstrahlung}
\label{sec:soft_real}
As the virtual corrections have an IR divergence, we also need to account for the soft bremsstrahlung. These correspond to diagrams, in which an additional soft photon is emitted from an external fermion line. Denoting the momentum of the fermion line with $l$ and the momentum of the soft photon by $k$, it corresponds to the amplitude:
\begin{equation}
    \mathcal{M}_s=\pm e Q_f\frac{\varepsilon^\ast \cdot l}{k\cdot l}\mathcal{M}_0,
\end{equation}
with the $+$ sign, if the fermion is outgoing and the $-$ sign if it is incoming, where $Q_f$ denotes the charge of the lepton, and where $\mathcal{M}_0$ denotes the amplitude without photon emission.  

To evaluate the soft bremsstrahlung contribution to the cross section, one has to integrate over the unobserved soft-photon phase space. It is easiest to perform this integral in a reference frame where the maximum soft-photon energy $\Delta E_s$ is isotropic. Such a reference frame depends on the specific experimental conditions. It is in general given by the rest frame of the soft photon and the unobserved particle in the process.

In Ref. \cite{Heller:2018ypa} we considered the case when the final proton is measured and the di-lepton pair remains undetected. Therefore, the integral has to be performed in the rest frame of the di-lepton pair and the soft photon, i.e. $\vec{l}_+ +\vec{l}_- +\vec{k}=0$. 

In this work, we consider the $\gamma p \to l^- l^+ p$ process where $l^-$ and $l^+$ are observed, while the recoiling nucleon remains unobserved. Therefore the bremsstrahlung contribution to the cross section  is evaluated 
in the rest frame of the unobserved proton and soft photon. Defining a missing momentum $p_m \equiv p^\prime + k$, this frame is defined by $\vec p_m = 0$. In such frame the bremsstrahlung contribution due to the soft-photon emission from the $l^-$ and $l^+$ is given by:
\begin{align}
 d\sigma_{s;r}=&-d\sigma_0\frac{e^2}{(2\pi)^3}\int_{|\vec{k}|<\Delta E_s}\frac{d^3\vec{k}}{2k^0}\left\{\frac{m^2}{(k \cdot l_+)^2}+\frac{m^2}{(k \cdot l_-)^2}\right.\nonumber\\
 &\left.-\frac{2(l_+ \cdot l_-)}{(k \cdot l_+)(k \cdot l_-)}\right\},\label{eq:softphotonint}
\end{align}
where the maximal soft-photon energy in this frame is denoted by $\Delta E_s$. 
We can easily perform the integrations for the first two terms in Eq.~(\ref{eq:softphotonint}), which yields the expression:
\begin{eqnarray}
 d\sigma_{s;r}&=&d\sigma_0 \left(\frac{- \alpha}{\pi}\right)
 \left\{ - \left[\frac{1}{\epsilon_\text{IR}}-\gamma_E+\ln\left(\frac{4\pi\mu^2}{4 (\Delta E_s)^2}\right)\right] \right. \nonumber \\
 && + \frac{1}{2 \tilde \beta_-}\ln\left(\frac{1- \tilde \beta_-}{1+ \tilde \beta_-}\right)
 + \frac{1}{2 \tilde \beta_+}\ln\left(\frac{1- \tilde \beta_+}{1+ \tilde \beta_+}\right) \nonumber \\
 && \left. - \frac{1}{2 \pi} I_{-+} \right\},
 \label{eq:soft1}
\end{eqnarray}
where $\tilde \beta_-$, $\tilde \beta_+$ are the lepton velocities, defined in the frame $\vec p_m = 0$:
\begin{eqnarray}
\tilde \beta_\mp = \left(1 - m^2 / \tilde E_\mp^2 \right)^{1/2}, 
\end{eqnarray}
where $\tilde E_\mp$ are the corresponding lepton energies, which can be expressed as:
\begin{eqnarray}
\tilde E_\mp &=& \frac{p_m \cdot l_\mp}{\sqrt{p_m^2}} 
\approx \frac{1}{M} (q + p - q^\prime) \cdot l_\mp.
\label{eq:leptonkin}
\end{eqnarray}
Furthermore in Eq.~(\ref{eq:soft1}), 
the integral $I_{-+}$ is due to the interference between soft-photon emissions from the $l^-$ and $l^+$ lines. It has been worked out e.g. in Ref. \cite{tHooft:1978jhc} as:
\begin{align}
   I&_{-+} \equiv\int_{|\vec{k}|<\Delta E_s}\frac{d^3 \vec k}{k^0}\frac{l_- \cdot l_+}{(k \cdot l_- )(k \cdot l_+ )}=  4\pi\frac{\eta \, l_- \cdot l_+ }{  \left( \eta \,  l_- \right)^2-l_+^2}\nonumber\\
   &\times \biggl\{- \frac{ 1 }{ 2 } \ln\left( \frac{ \left( \eta l_- \right)^2 }{ l_+^2 } \right)   \left[ \frac{ 1 }{ \epsilon_{\mathrm{IR}} } - \gamma_E + \ln\left( \frac{ 4 \pi \mu^2 }{4 (\Delta E_s)^2 } \right)\right] \nonumber 
   \\&   + \biggl[ \frac{ 1 }{ 4 } \ln^2\left( \frac{ u_0 - \left\lvert \vec{u} \right\rvert}{   u_0 + \left\lvert \vec{u} \right\rvert} \right) + \text{Li}_2\left( 1-\frac{u_0+ \left\lvert \vec{u} \right\rvert  }{ v } \right) \nonumber\\
   &+ \text{Li}_2\left( 1-\frac{u_0-\left\lvert \vec{u} \right\rvert  }{ v } \right)\biggl]^{u=\eta l_-}_{u=l_+}  \biggl\},\label{eq:soft_real_Int}
\end{align}
\noindent
with
\begin{eqnarray}
    \eta&\equiv&\frac{l_- \cdot l_+}{m^2}
    +\sqrt{\left(\frac{l_- \cdot l_+}{m^2}\right)^2- 1 } = \frac{1 + \beta}{1 - \beta},  \\
    v&\equiv&\frac{ ( \eta^2 - 1) m^2 }{2 \left( \eta l_{-} - l_{+} \right)_0 } 
    = \frac{\beta s_{ll}}{2 (\tilde E_- - \frac{1 - \beta}{1 + \beta} \tilde E_+)},
\end{eqnarray}
with $\beta$ given in Eq.~\eqref{EqBeta}. 

In general we can divide the soft-photon contribution of Eq.~(\ref{eq:soft1}) in an IR divergent piece ($\delta_\text{s;r}^\text{IR}$) and a finite piece ($\delta_\text{s;R}$) as:  
\begin{align}
 d\sigma_{s;r}&= d\sigma_0\left(\vphantom{\frac{1}{2}}\delta_\text{s;r}^\text{IR}+\delta_\text{s;r}\right), 
\end{align}
with IR divergent piece given by:
\begin{align}
\delta^\text{IR}_\text{s;r}=&\left(\frac{-\alpha}{\pi}\right)\left[\left(\frac{1+\beta^2}{2\beta}\right)\ln\left(\frac{1+\beta}{1-\beta}\right)-1\right]\nonumber\\
&\times\left[\frac{1}{\epsilon_\text{IR}}-\gamma_E+\ln\left(\frac{4\pi\mu^2}{m^2}\right)\right], \label{eq:IR_real}
\end{align}
and the finite part $\delta_\text{s;r}$ expressed as:
\begin{align}
\delta_\text{s;r}=&\left(\frac{-\alpha}{\pi}\right)\left\{\ln\left(\frac{4 (\Delta E_s)^2}{m^2}\right)\left[1-\left(\frac{1+\beta^2}{2\beta}\right)\ln\left(\frac{1+\beta}{1-\beta}\right)\right]\right.\nonumber\\
&+ \frac{1}{2 \tilde \beta_-}\ln\left(\frac{1- \tilde \beta_-}{1+ \tilde \beta_-}\right)
 + \frac{1}{2 \tilde \beta_+}\ln\left(\frac{1- \tilde \beta_+}{1+ \tilde \beta_+}\right)
 \nonumber \\
 &- \left(\frac{1+\beta^2}{2\beta}\right)
 \left[ \frac{1}{4} \ln^2 \left(\frac{1-\tilde \beta_-}{1+ \tilde \beta_-}\right)
 - \frac{1}{4} \ln^2 \left(\frac{1-\tilde \beta_+}{1+ \tilde \beta_+}\right)
 \right. \nonumber \\
 &+ \text{Li}_2\left(1 - \left(\frac{1+\beta}{1-\beta}\right) \frac{\tilde E_-}{v} (1 + \tilde \beta_-) \right) \nonumber \\
 &+ \text{Li}_2\left(1 - \left(\frac{1+\beta}{1-\beta}\right) \frac{\tilde E_-}{v} (1 - \tilde \beta_-) \right) \nonumber \\
  &- \text{Li}_2\left(1 - \frac{\tilde E_+}{v} (1 + \tilde \beta_+) \right)  \nonumber \\
 &\left. \left. - \text{Li}_2\left(1 - \frac{\tilde E_+}{v} (1 - \tilde \beta_+) \right) 
 \right] \right\}. 
\label{eq:finite_real}
\end{align}
In the ultrarelativistic limit for the leptons, i.e. $\beta \approx 1$, $\tilde \beta_\mp \approx 1$, which is a very good approximation for the production of an $e^- e^+$ pair as considered in the following, the above expression simplifies  considerably. Using 
\begin{eqnarray}
v \approx s_{ll} / (2 \tilde E_-),
\end{eqnarray}
we obtain in this limit for the finite part $\delta_\text{s;r}$:
\begin{eqnarray}
\delta_\text{s;r}&\approx&\left(\frac{\alpha}{\pi}\right)\left\{\ln\left(\frac{(\Delta E_s)^2}{ \tilde E_- \tilde E_+}\right)\left[\ln\left(\frac{s_{ll}}{m^2}\right) - 1\right]
+ \frac{1}{2}\ln^2 \left(\frac{s_{ll}}{m^2}\right) \right.\nonumber\\
&-&\left. \frac{1}{2}  \ln^2 \left( \frac{\tilde E_-}{\tilde E_+} \right)  
 -  \frac{\pi^2}{3} +  
 \text{Li}_2\left(1 - \frac{s_{ll}}{4 \tilde E_- \tilde E_+} \right)  \right\}. 
\label{eq:finiterealultrarel}
\end{eqnarray}
Note that the IR divergent pieces from virtual and real corrections, ie. Eqs. \eqref{eq:soft_div_virt} and \eqref{eq:IR_real}, exactly cancel on the level of the cross section, thus giving an IR finte result.

For the purpose of estimating the di-lepton forward-backward asymmetry in the following, it will be convenient to express the lepton angles in the $l^- l^+$ rest frame as defined in \cite{Pauk:2020gjv}, which are denoted by $\theta_l, \phi_l$ for polar and azimuthal angles respectively. 
Eq.~(\ref{eq:leptonkin}) then allows to express the di-lepton energies $\tilde E_{\mp}$ in terms of invariants and the angle $\theta_l$ as:
\begin{eqnarray}
\tilde E_{\mp} &=& \frac{1}{4 M} \left\{ 
 (W^2 - M^2 - s_{ll})
\right. \nonumber \\
&\pm&\left.  
[(W^2 - M^2 - s_{ll})^2 - 4 M^2 s_{ll}]^{1/2}
\beta \cos \theta_l \right\}.
\end{eqnarray}

To evaluate the finite soft-photon cross section corrections of Eq.~(\ref{eq:finite_real}) or Eq.~(\ref{eq:finiterealultrarel}), we need to express the maximal soft-photon energy $\Delta E_s$, defined in the frame $\vec p_m = 0$ in terms of the experimental resolutions using:  
\begin{eqnarray}
\Delta E_s = \Delta\left( \frac{p_m^2 - M^2}{2 \sqrt{p_m^2}} \right) \approx  \frac{ \Delta p_m^2}{2 M},
\end{eqnarray}
where to first order we have used $p_m^2 \approx M^2$ in the denominator, and where $\Delta p_m^2$ denotes the resolution in the missing mass squared.  For the case of detecting the di-lepton pair, the lab energies ($E_\mp$), the scattering angle between the pair ($\theta^\text{Lab}_{ll}$), and the angle between the incoming photon $q$ and the virtual photon $q^\prime$ ($\theta^{Lab}_{\gamma\gamma}$)  are measured.  These four parameters are in correspondence with the kinematic quantities  $s_{ll}$, $t$, $\theta_l$, and $\phi_l$ introduced in the cross section expression of Eq.~(\ref{CrossSectionSum}). 
 
In the Lab frame, the missing mass squared can be expressed, neglecting the lepton mass in the kinematics, as:
\begin{eqnarray}
p_m^2 &=&(q-q'+p)^2 \\
&=& M^2 - 2 q \cdot q' + s_{ll} + 2 p \cdot (q - q') \nonumber \\
&=& \Big[ M^2+ 2 M E_\gamma + 2 E_- E_+ \left( 1 - \cos\theta_{ll}\right)
 \nonumber \\
&+&2 E_\gamma\lvert\pvec{q}'\rvert \cos\theta_{\gamma\gamma}   
-2 (E_\gamma + M) (E_+ + E_-) \Big]_{\text{Lab}}.
\nonumber
\end{eqnarray}
Accounting for the finite resolutions in the Lab kinematical quantities, and adding their contributions quadratically, we can express the maximal soft-photon energy $\Delta E_s$ as:
\begin{eqnarray}
    \Delta E_s &=& \frac{1}{M} \Big[ 
    \Big(- (E_\gamma + M) +  E_+ (1 - \cos\theta_{ll})   \nonumber \\
    && \quad + (E_- + E_+ \cos\theta_{ll}) \frac{E_\gamma}{\lvert\pvec{q}'\rvert}\cos \theta_{\gamma\gamma} \Big)^2 (\Delta E_-)^2  \nonumber\\
    &&\quad + \Big(- (E_\gamma + M) +  E_- (1 - \cos\theta_{ll})   \nonumber \\
    && \quad + (E_+ + E_- \cos\theta_{ll}) \frac{E_\gamma}{\lvert\pvec{q}'\rvert}\cos \theta_{\gamma\gamma} \Big)^2 (\Delta E_+)^2  \nonumber\\
    &&\quad + E_-^2 E_+^2 (1 - \frac{E_\gamma}{\lvert\pvec{q}'\rvert} \cos \theta_{\gamma\gamma})^2 \sin^2 \theta_{ll}  (\Delta \theta_{ll})^2 \nonumber \\
    && \quad + E_\gamma^2 \lvert\pvec{q}'\rvert^2 \sin^2 \theta_{\gamma\gamma} (\Delta \theta_{\gamma\gamma})^2 \Big]^{1/2}_{\text{Lab}}.
    \label{eq:deltasoft}
\end{eqnarray}

As is evident from Eq.~(\ref{eq:deltasoft}), the evaluation of the soft-photon radiative correction depends on the specific experimental resolutions. For the purpose of illustrating the size of these corrections, we will provide predictions below where the soft-photon cut-off energy $\Delta E_s$ is chosen to be in the one to few percent range of the beam energy, as a realistic value. 
For our predictions in the low-energy region ($E_\gamma \simeq 0.36$~GeV), we will show the corrections for $\Delta E_s = 0.01$~GeV, whereas for the high-energy region ($E_\gamma \simeq 6.8$~GeV), we will show the results for $\Delta E_s = 0.05$~GeV.

\section{Results and discussion\label{sec:results}}

\subsection{Observables\label{sec:result_observable}}

We use our setup to study the radiative corrections 
to the $\gamma p \to e^-e^+ p$ process in both low- and high-energy regimes. For both kinematical situations we study the effect of these corrections on the cross section, on the forward-backward asymmetry $A_{FB}$, as well as on the beam helicity asymmetry $A_{\odot}$. These asymmetries are respectively defined as
\begin{align}
    A_{FB}&=\frac{d\sigma_{\theta_l,\phi_l}-d\sigma_{\pi - \theta_l,\phi_l + \pi}}{d\sigma_{\theta_l,\phi_l}+d\sigma_{\pi - \theta_l,\phi_l + \pi}},
    \label{eq:asymmfb}\\
    A_{\odot}&=\frac{d\sigma^+-d\sigma^-}{d\sigma^+ +d\sigma^-},\label{eq:asymmhel}
\end{align}
where $d\sigma_{\theta_l,\phi_l}$ in $A_{FB}$ stands for the unpolarized cross section measured at lepton angles $\theta_l$ and $\phi_l$ (in the $l^-l^+$ rest frame), 
and where $d\sigma^{\pm}$ in $A_{\odot}$ stand for the polarized cross sections for circular photon polarization $\pm 1$ respectively.  

Because of the opposite symmetry of the BH and TCS amplitudes 
(Fig.~\ref{fig:tree}) under charge conjugation (odd versus even number of photon couplings to the lepton charge),
 the asymmetry $A_{FB}$, which interchanges the kinematics for $l^-$ and $l^+$, allows for a direct assessment of the interference term
between the BH and the real part of the TCS amplitude.  
The BH and TCS processes separately yield a zero asymmetry.
While $A_{FB}$ is proportional to the real part of the BH-TCS interference, $A_\odot$ is proportional to the imaginary part of this interference.  
Note however that the complex TCS amplitude by itself also yields a contribution to $A_\odot$, which is usually very small unless the real part of the TCS amplitude becomes comparable in size to the BH amplitude. The observables $A_{FB}$ and $A_{\odot}$ are thus complementary in accessing the complex TCS amplitude experimentally and in testing theoretical models.

In order to calculate the TCS observables we implemented all amplitudes in a \texttt{C++} code in which the interference of different Feynman amplitudes can be evaluated numerically.

\subsection{Results for TCS observables in the $\Delta(1232)$ region}

We firstly study the importance of the radiative corrections on the $\gamma p \to e^-e^+ p$ observables in the low-energy  kinematical region. The $\gamma p \to e^-e^+ p$ process was studied in Ref.~\cite{Pauk:2020gjv} in the $\Delta(1232)$ resonance region at small values of $s_{ll}$ and $-t$. 
In this limit, the TCS amplitude approaches the forward real Compton scattering amplitude, for which a full dispersive calculation based on empirical structure functions exists~\cite{Gryniuk:2015eza}. It was found in~\cite{Pauk:2020gjv} that around 
$W = 1.25$~GeV the full TCS cross section integrated over the di-lepton angles is reproduced by a Born + $\Delta$-pole model for the TCS amplitude, as discussed in Sec.~\ref{sec:dvcs_low_energy}, within an accuracy of 5\% or better. Therefore, we consider the Born + $\Delta(1232)$-pole model to be realistic enough as a model for the TCS amplitude around the $\Delta(1232)$-pole in the near forward direction in order to study the effect of the QED radiative corrections on this reaction.   
In the following, we will show the effect of the radiative corrections on the $\gamma p \to e^-e^+ p$ cross section as well as on the two asymmetries $A_{FB}$ and $A_\odot$ for a c.m. energy of $W=1.25$ GeV. 

\begin{figure}
	\centering
 \includegraphics[scale=0.7]{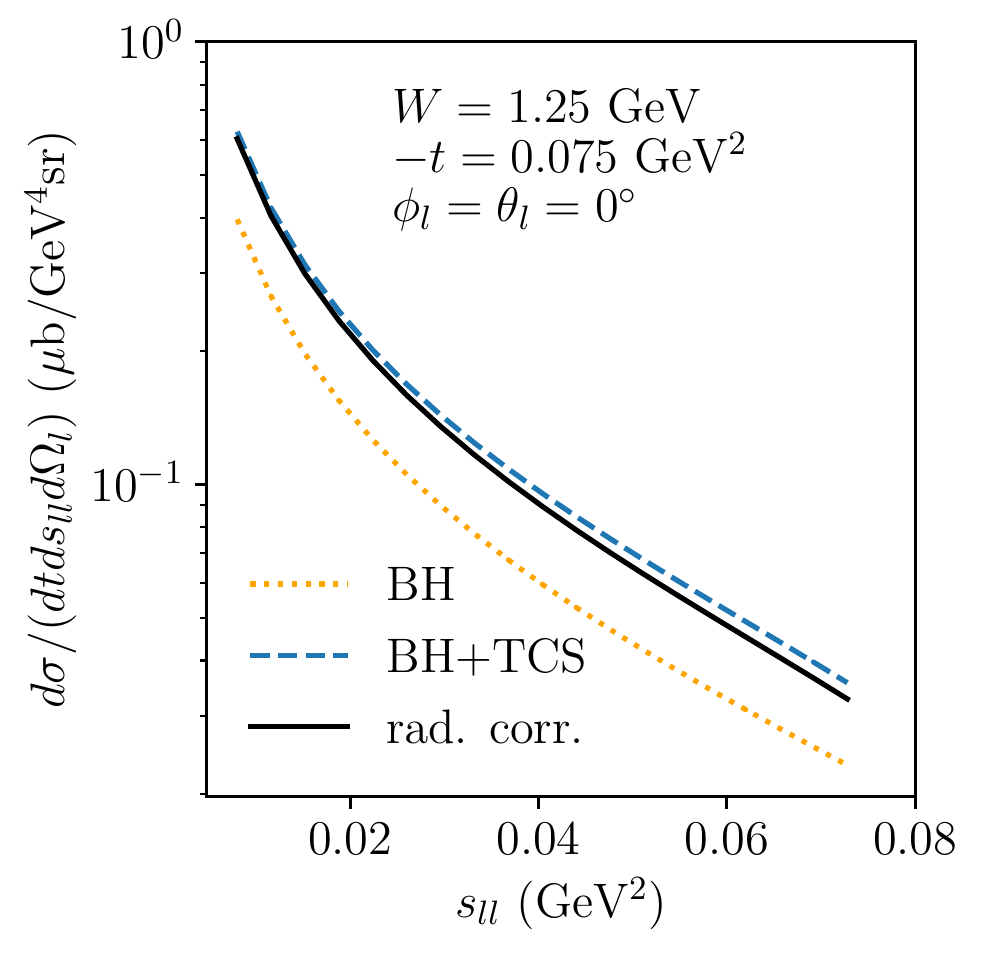}
 \includegraphics[scale=0.7]{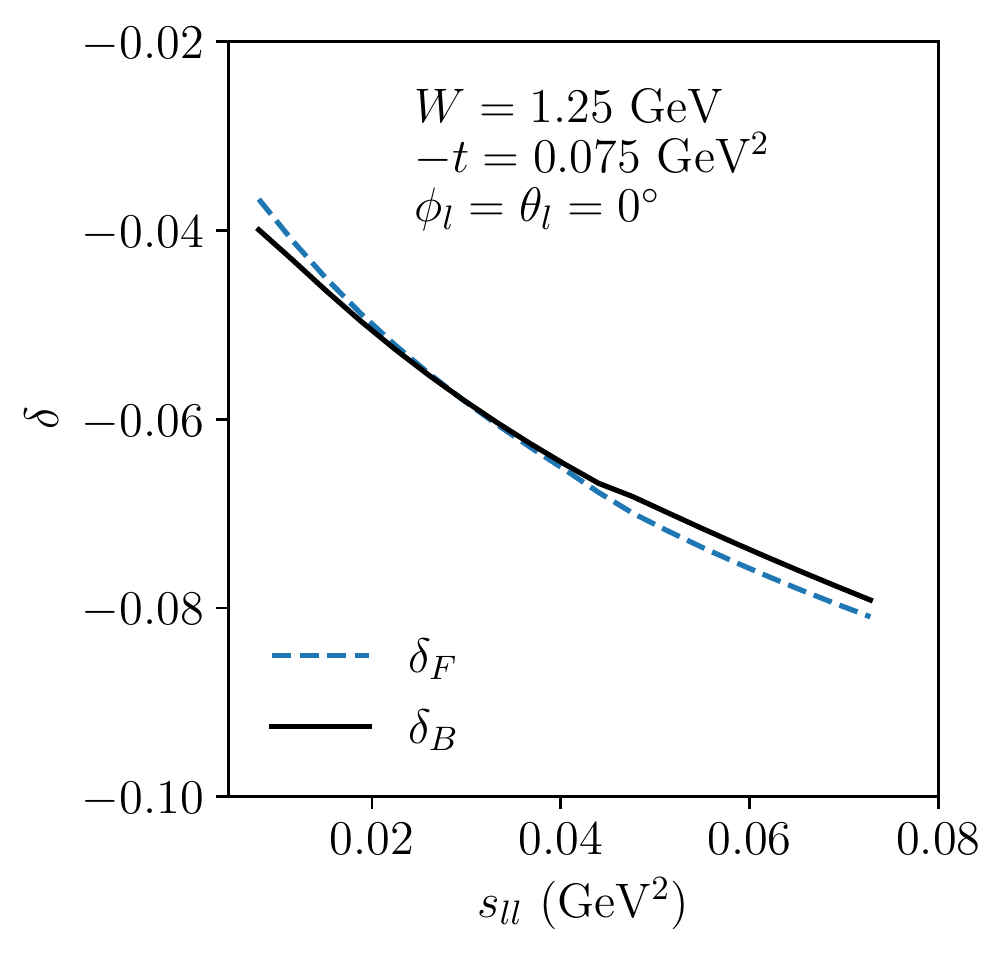}
  \includegraphics[scale=0.7]{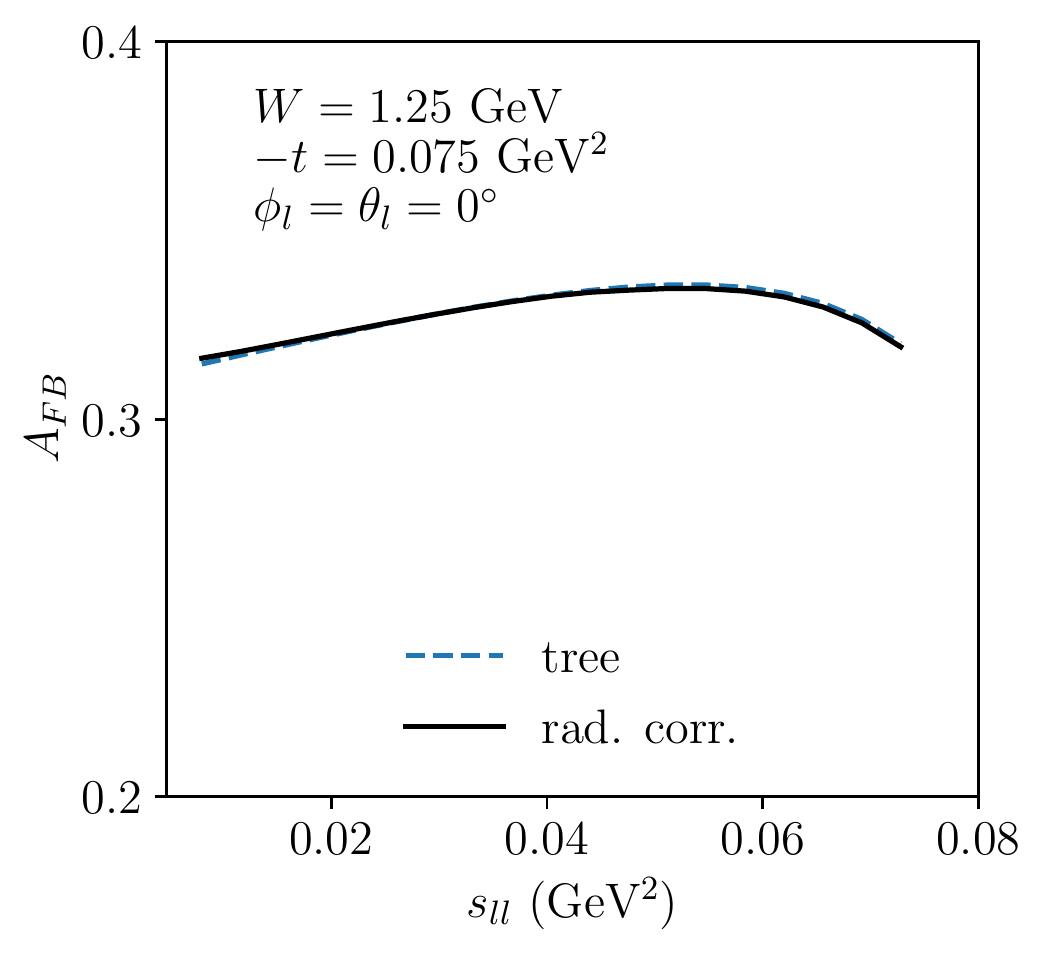}
	\caption{Upper panel: $s_{ll}$-dependence of the $\gamma p \to e^-e^+ p$ unpolarized cross section for in-plane kinematics of the di-lepton pair. We show results for BH, for BH+TCS using a Born + $\Delta(1232)$-pole model for TCS, and when including the first-order QED radiative corrections to BH+TCS, for a soft-photon energy of $\Delta E_s=0.01$ GeV. The middle panel displays the radiative correction factor for the forward ($\delta_F$) and backward ($\delta_B$) cross sections. The lower panel shows the sensitivity of the forward-backward asymmetry $A_{FB}$ on the TCS amplitude, and the negligible radiative correction on this observable (solid and dashed curves nearly overlapping). \label{fig:low_energy_Afb}}
\end{figure}

\begin{figure}
	\centering
 \includegraphics[scale=0.7]{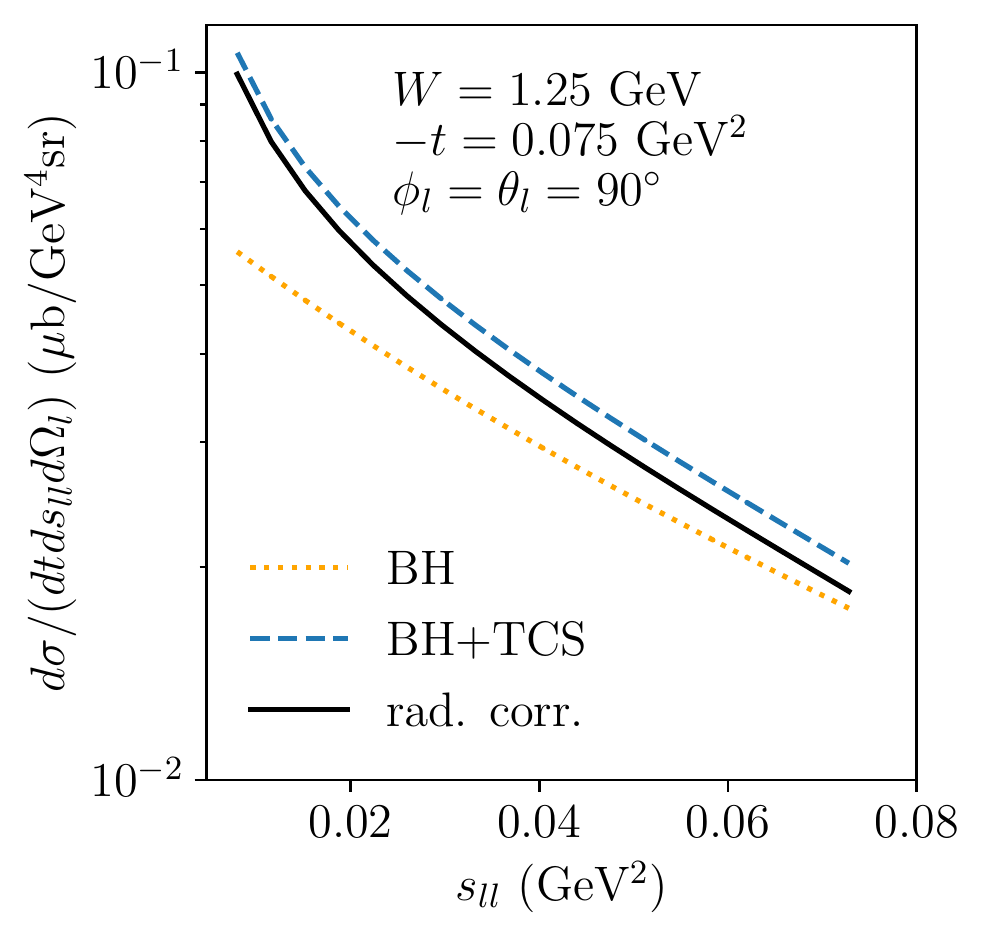}
 \includegraphics[scale=0.7]{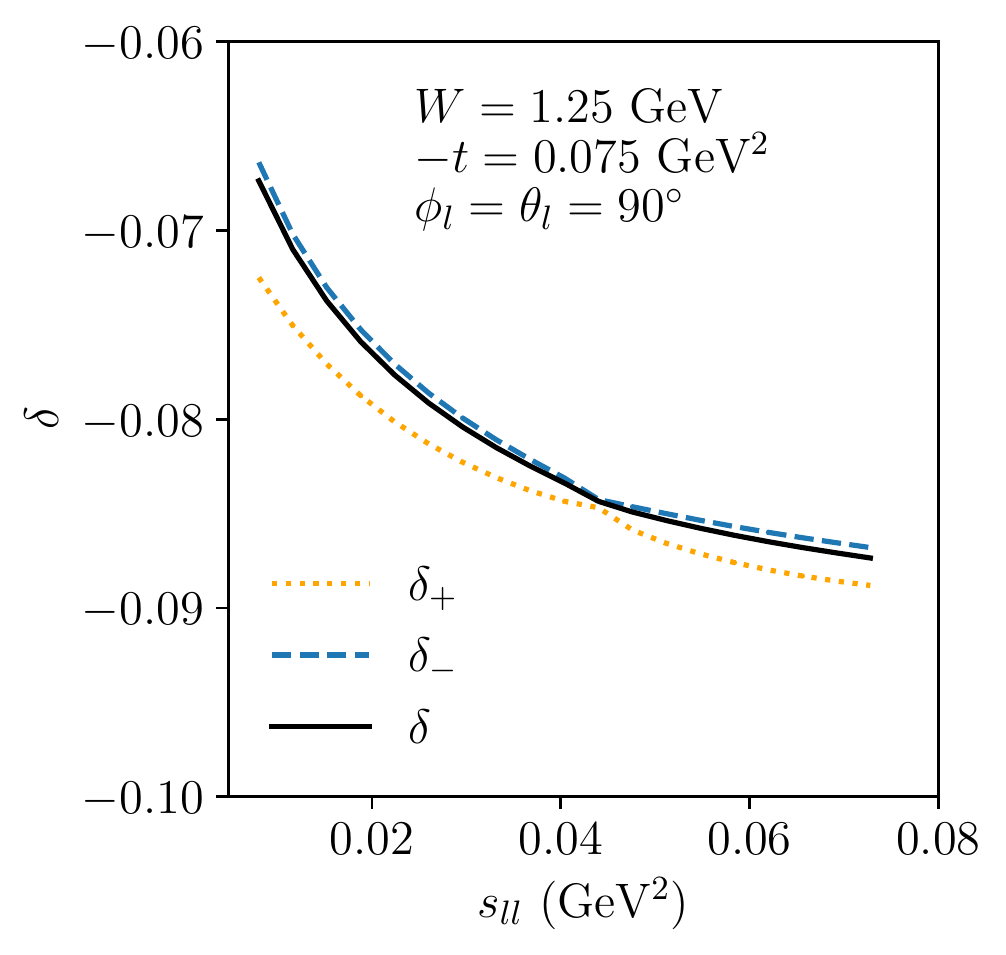}
  \includegraphics[scale=0.7]{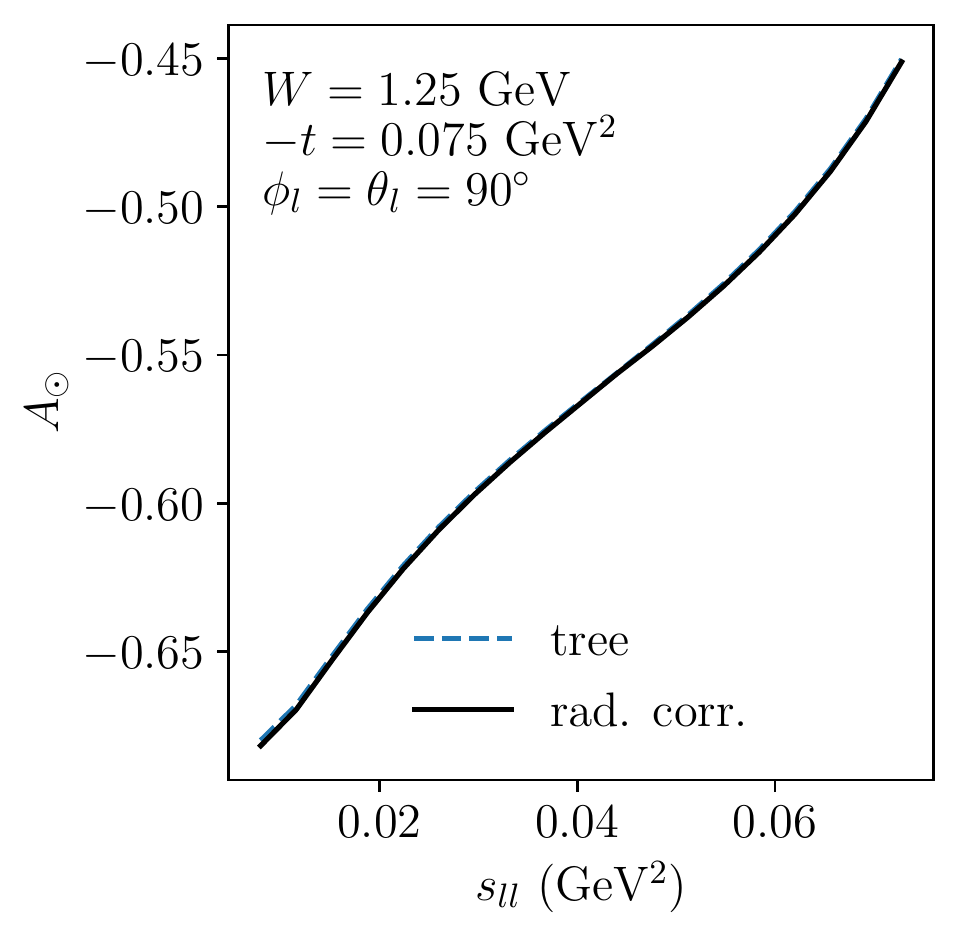}
	\caption{Upper panel: $s_{ll}$-dependence of the $\gamma p \to e^-e^+ p$ unpolarized cross section for out-of-plane kinematics of the di-lepton pair. We show results for BH, for BH+TCS using a Born + $\Delta(1232)$-pole model for TCS, and when including the first-order QED radiative corrections to BH+TCS, for a soft-photon energy of $\Delta E_s=0.01$ GeV. 
	The middle panel shows the cross section correction factors $\delta_\pm$ for circular photon polarization $\pm 1$. The lower panel shows the sensitivity of the photon beam helicity asymmetry $A_{\odot}$ on the TCS amplitude, and the negligible radiative correction on this observable. \label{fig:low_energy_A_hel}}
\end{figure}

In Fig. \ref{fig:low_energy_Afb} we show the $s_{ll}$-  dependence of the  unpolarized $\gamma p \to e^-e^+ p$ observables for in-plane di-lepton kinematics ($\theta_l = 0^\circ$ and $\phi_l = 0^\circ$) in the $\Delta(1232)$ region and for a small value of $-t$. One notices that the TCS amplitude increases the cross section in this kinematical region by 50\% or more, and that the forward-backward asymmetry $A_{FB}$, which depends linearly on the TCS amplitude reaches values larger than 30\%. 
We see that the first-order QED radiative corrections yield a 5 - 8\% correction on the $\gamma p \to e^-e^+ p$ cross section in this kinematical range. However, we also notice that in comparing the complementary forward ($\delta_F$ for $\theta_l=\phi_l=0^{\circ}$) and backward ($\delta_B$ for $\theta_l=\phi_l=180^{\circ}$) kinematics, these correction factors are nearly the same, such that the asymmetry $A_{FB}$ is to good approximation unaffected by the first-order QED radiative corrections. This makes $A_{FB}$ an ideal observable to extract the real part of the TCS amplitude.  

Until recently, the only proof-of-principle experiment of 
the forward-backward asymmetry of the $\gamma p \to e^- e^+ p$ process was conducted by Alvensleben {\it et al.} at DESY~\cite{Alvensleben:1973mi}. It was performed at $W \simeq 2.2$~GeV in near-forward kinematics, and aimed at an experimental verification of the Kramers-Kronig dispersion relation for the forward Compton amplitude. 
The Kramers-Kronig relation encompasses a fundamental connection  between the photon absorption and scattering based on analyticity and unitarity, and allows to evaluate the real part of the forward Compton scattering off protons through a dispersive integral over its imaginary part, which is evaluated using the empirical knowledge of the total photoabsorption cross sections. It was shown in a recent re-analysis of the Kramers-Kronig relation for forward Compton scattering~\cite{Gryniuk:2015eza} that the present database of the unpolarized photoabsorption cross section is not entirely consistent, and shows the largest discrepancies (in the 5 - 10\% range) in the region of the $\Delta(1232)$ peak and in its higher-energy tail region. It was furthermore shown in \cite{Gryniuk:2015eza} that these discrepancies in the world database also directly limit the obtained precision of the extracted Baldin sum rule value for the sum of proton electric and magnetic polarizabilities, $\alpha_{E1} + \beta_{M1}$. 
 Being nearly unaffected by the first-order QED radiative corrections, our results show that the forward-backward asymmetry in the resonance region provides a direct measurement of the real part of the near-forward Compton amplitude and thus has the potential to settle the existing discrepancies in the Compton database.

In Fig. \ref{fig:low_energy_A_hel} we show the 
corresponding di-lepton mass dependence of the $\gamma p \to e^-e^+ p$ observables for a circularly polarized photon beam for out-of-plane di-lepton kinematics ($\theta_l = \phi_l = 90^\circ$). As before we see that the first-order QED radiative corrections yield a 7 - 9\% correction on the $\gamma p \to e^-e^+ p$ cross section in this kinematical range. The small cusp in the curves of $\delta^{\pm}$ around $s_{ll}\approx 0.044$ GeV$^2$ is due to the muon threshold in the vacuum polarization. This gives a contribution with different sign for the cross sections $d\sigma ^{\pm}$.
For the corresponding photon helicity asymmetry $A_\odot$, which is directly proportional to the imaginary part of the BH-TCS interference, the first-order QED radiative corrections nearly drop out.
For kinematics close to the $\Delta(1232)$-pole position, where the imaginary part of the TCS amplitude is maximal, we find a large value of the asymmetry, varying between $-65\%$ and $-45\%$. 

\begin{figure}
	\centering
 \includegraphics[scale=0.6]{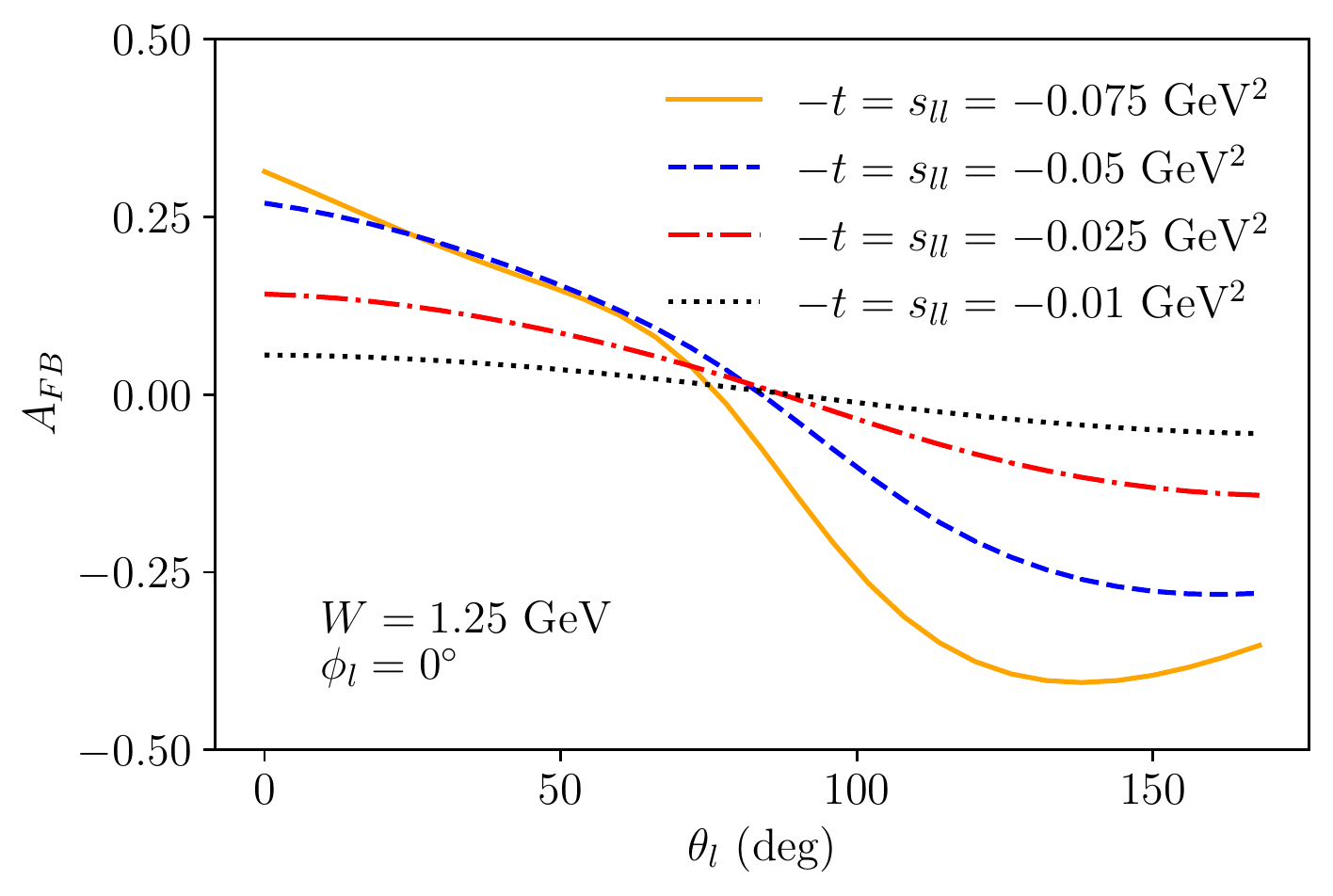}
 \includegraphics[scale=0.6]{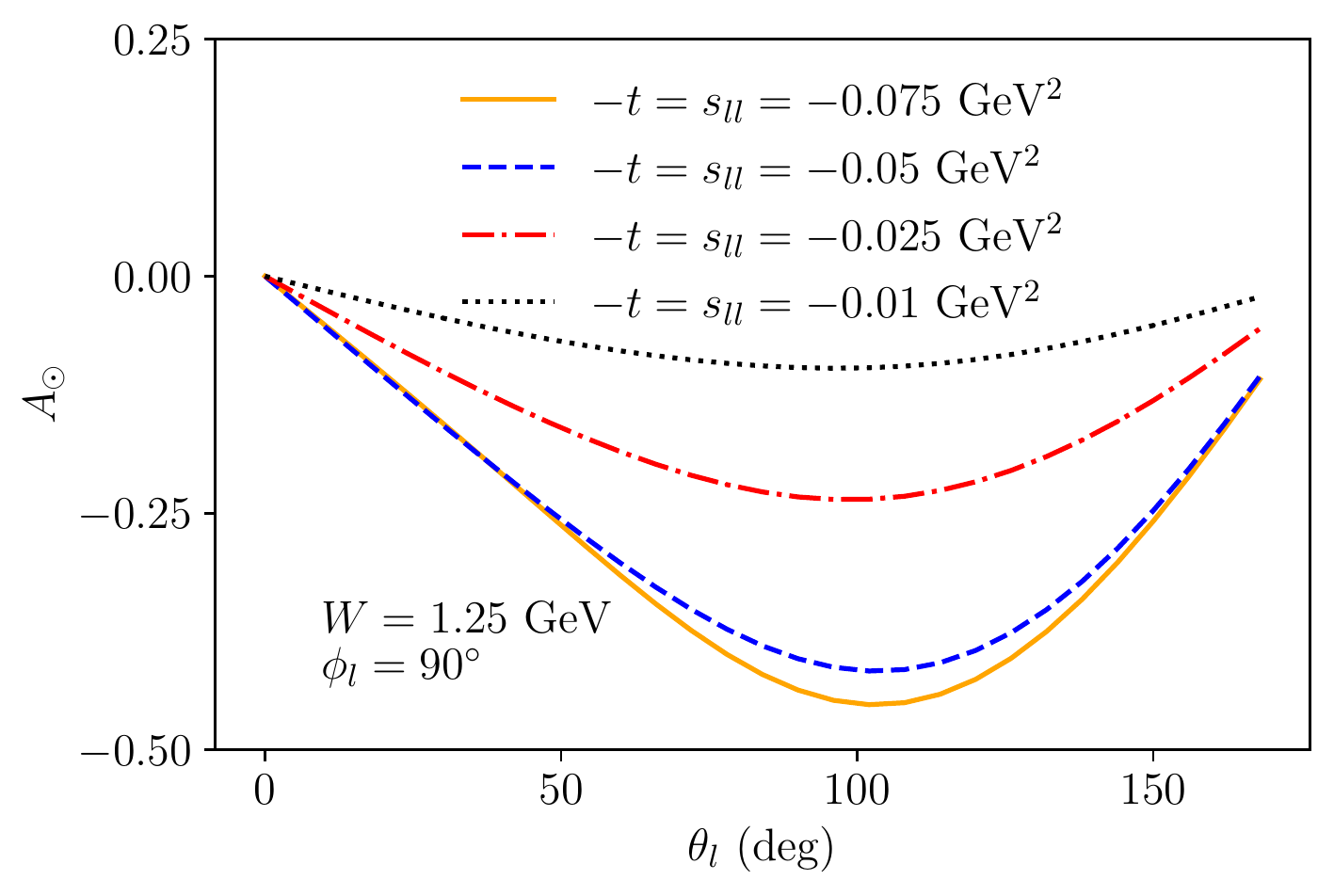}
	\caption{Dependence of the forward-backward asymmetry (upper panel) and photon beam helicity asymmetry (lower panel) on the lepton angle $\theta_l$ for different values of $s_{ll}$ and $t$. The effect of the radiative corrections on both observables is negligibly small. 
	\label{fig:low_energy_thetal}}
\end{figure}

In Fig.~\ref{fig:low_energy_thetal} we show the dependence of both asymmetries in the $\Delta(1232)$ region on the lepton angle $\theta_l$ for fixed out-of-plane angle $\phi_l$ ($\phi_l=0^{\circ}$ for $A_{FB}$ and $\phi_l=90^{\circ}$ for $A_\odot$) and different values of $t$ and $s_{ll}$. We see that increasing the values of $-t$ and $s_{ll}$ results in larger asymmetries due to the larger interference terms. We only show the radiatively corrected results in Fig.~\ref{fig:low_energy_thetal}, as the difference between the tree-level result and the result including first-order QED radiative corrections is at the few per mille level on these  asymmetries.

\subsection{Results for high-energy TCS observables}

In this section we study the effect of the first-order QED radiative corrections on the cross section as well as on the forward-backward and beam helicity asymmetries of the $\gamma p \to e^-e^+ p$ process in the high-energy, forward scattering regime in which the di-lepton pair is produced with a large virtuality. In this regime, a QCD factorization theorem allows to model the TCS amplitude in terms of GPDs as described in Sec. \ref{sec:dvcs_high_energy}. We present here results in the kinematics of a recent CLAS12@JLab experiment~\cite{JLabtcs:2020}. The latter is the first TCS experiment where data have been reported~\cite{Chatagnon:2020egu,Chatagnon:2020} in the kinematical regime  corresponding with an average c.m. energy of $W=3.69$ GeV and a large di-lepton invariant mass squared of $s_{ll}=3.24$ GeV$^2$. 

\begin{figure}
	\centering
 \includegraphics[scale=0.7]{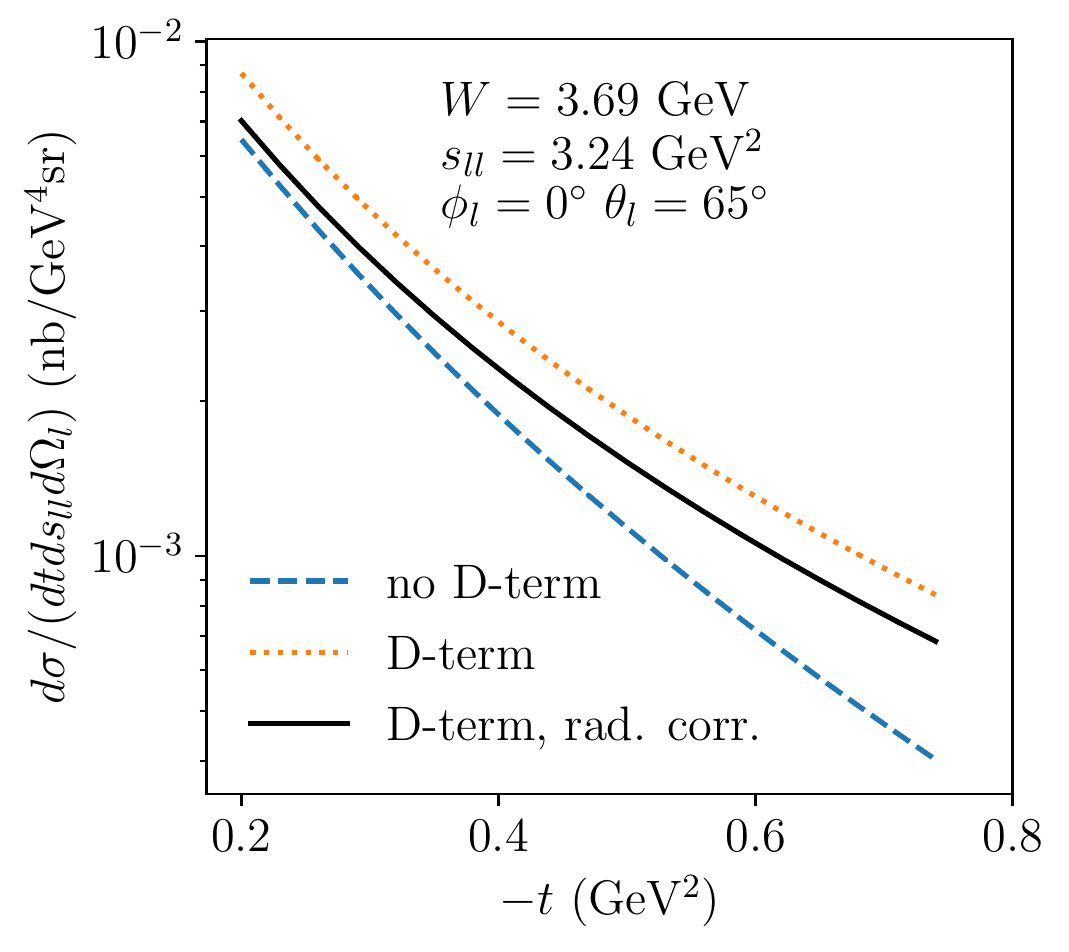}
 \includegraphics[scale=0.7]{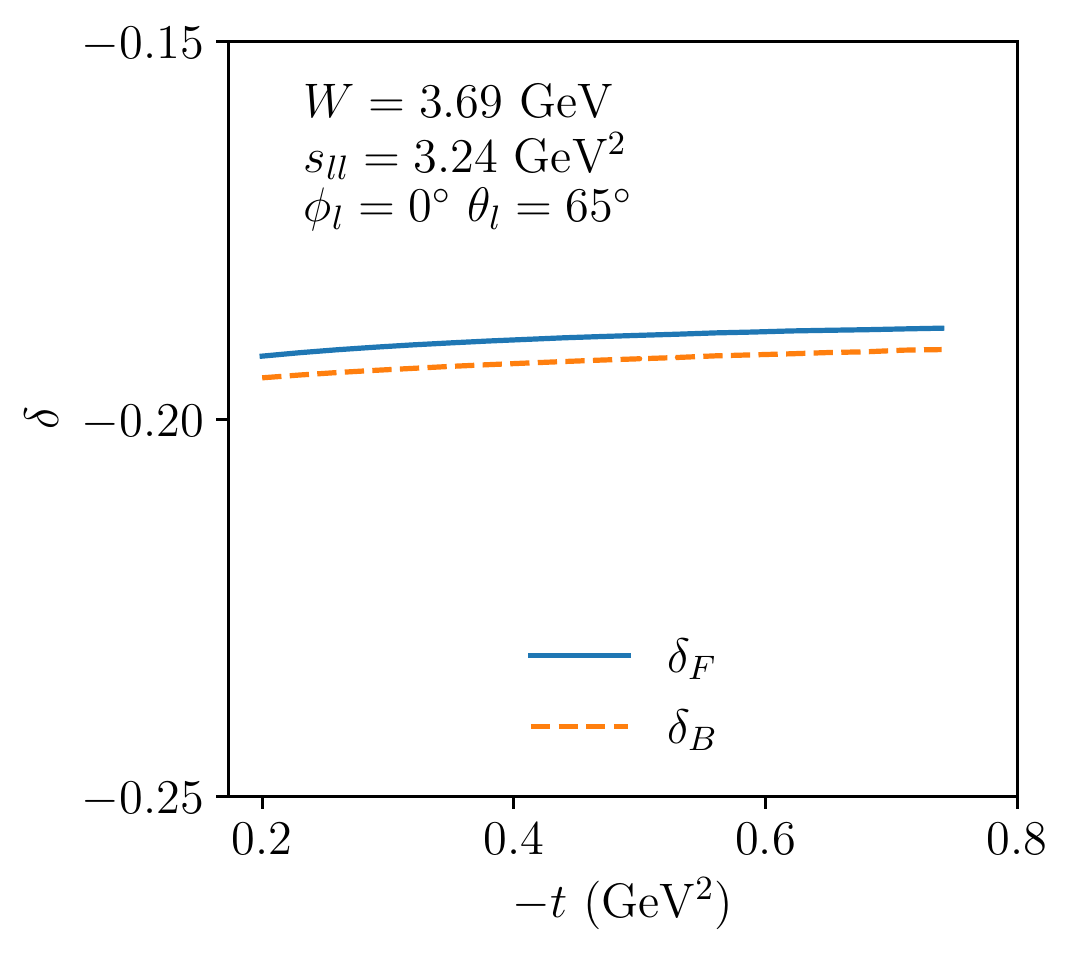}
  \includegraphics[scale=0.7]{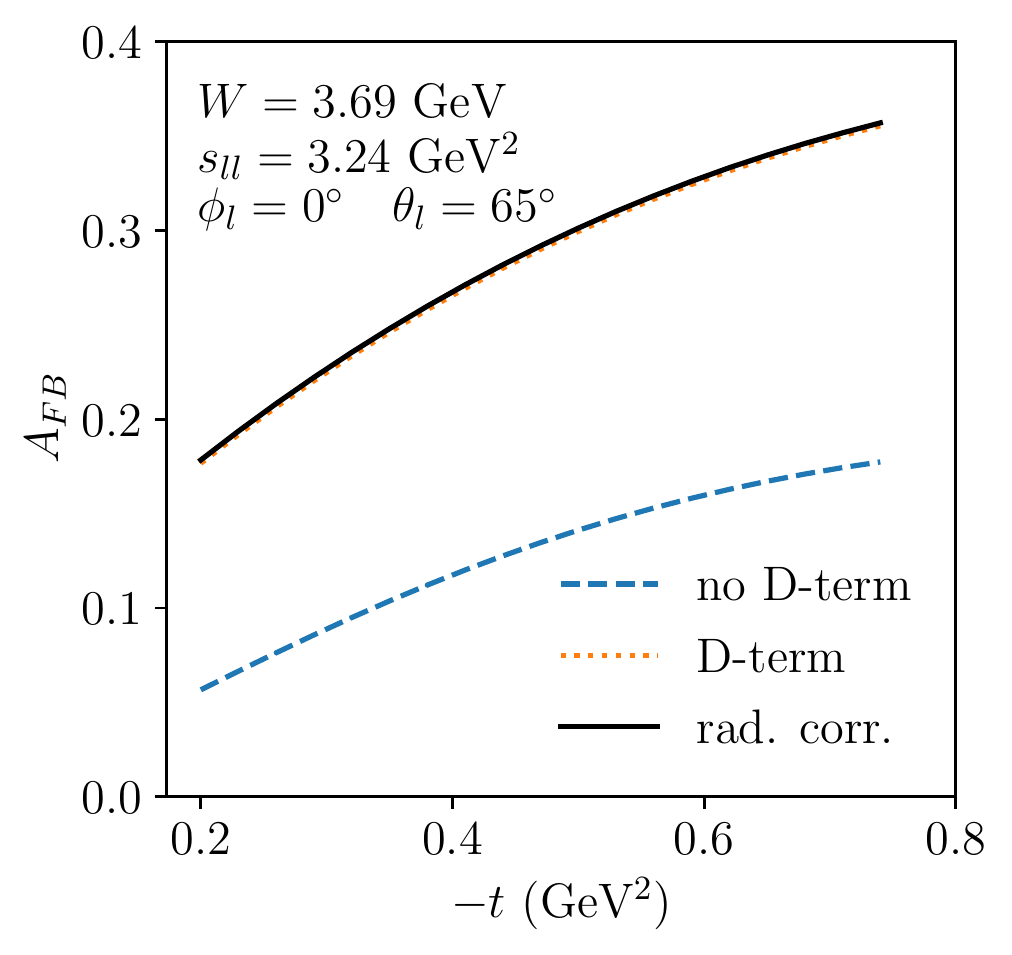}
	\caption{Upper panel: $t$-dependence of the $\gamma p \to e^-e^+ p$ unpolarized cross section for in-plane kinematics of the di-lepton pair corresponding with the CLAS12@JLab experiment~\cite{Chatagnon:2020egu,Chatagnon:2020}. We show our results for two GPD models and including the first-order QED radiative corrections, for a soft-photon energy of $\Delta E_s=0.05$ GeV. The middle panel displays the radiative correction factor for the forward ($\delta_F$ for $\theta_l = 65^{\circ}$, $\phi_l = 0^{\circ}$) and backward ($\delta_B$ for $\theta_l = 115^{\circ}$, $\phi_l = 180^{\circ}$) cross sections. The lower panel shows the sensitivity of the forward-backward asymmetry $A_{FB}$ on the D-term contribution to the GPD, and the negligible radiative correction on this observable (solid and dotted curves nearly overlapping). 
	\label{fig:high_energy_AFB}}
\end{figure}

In Fig. \ref{fig:high_energy_AFB} we show the $t$-dependence of the  unpolarized $\gamma p \to e^-e^+ p$ observables for in-plane di-lepton kinematics ($\theta_l = 65^\circ$ and $\phi_l = 0^\circ$), corresponding with the CLAS12@JLab experimental conditions. One notices that the cross section and the forward-backward asymmetry show a sizeable sensitivity on the D-term contribution to the GPD parameterization, which contributes to the real part of the TCS amplitude. Comparing the GPD double distribution parameterization (shown by blue dashed curves in upper and lower panels in Fig.~\ref{fig:high_energy_AFB}), which was used in a previous global analysis of DVCS data in the valence region~\cite{Dupre:2016mai,Dupre:2017hfs}, we see that adding the dispersive estimate of Ref.~\cite{Pasquini:2014vua} for the D-term contribution (dotted red curves) gives a sizeable contribution to the cross section. For the forward-backward asymmetry, which is directly proportional to the real part of the BH-TCS interference, it leads to a shift in the asymmetry by 10 - 20\%. We see that the first-order QED radiative corrections yield a nearly 20\% correction on the $\gamma p \to e^-e^+ p$ cross section, which are important to account for in the extraction of the Compton form factors. However, we also notice that in comparing the complementary forward and backward kinematics, these correction factors are nearly the same, such that the asymmetry $A_{FB}$ is to good approximation unaffected by the first-order QED radiative corrections. This makes $A_{FB}$ a gold-plated observable to extract the real part of the Compton form factor, and test its sensitivity to the D-term contribution in GPD parameterizations. 

\begin{figure}
	\centering
 \includegraphics[scale=0.7]{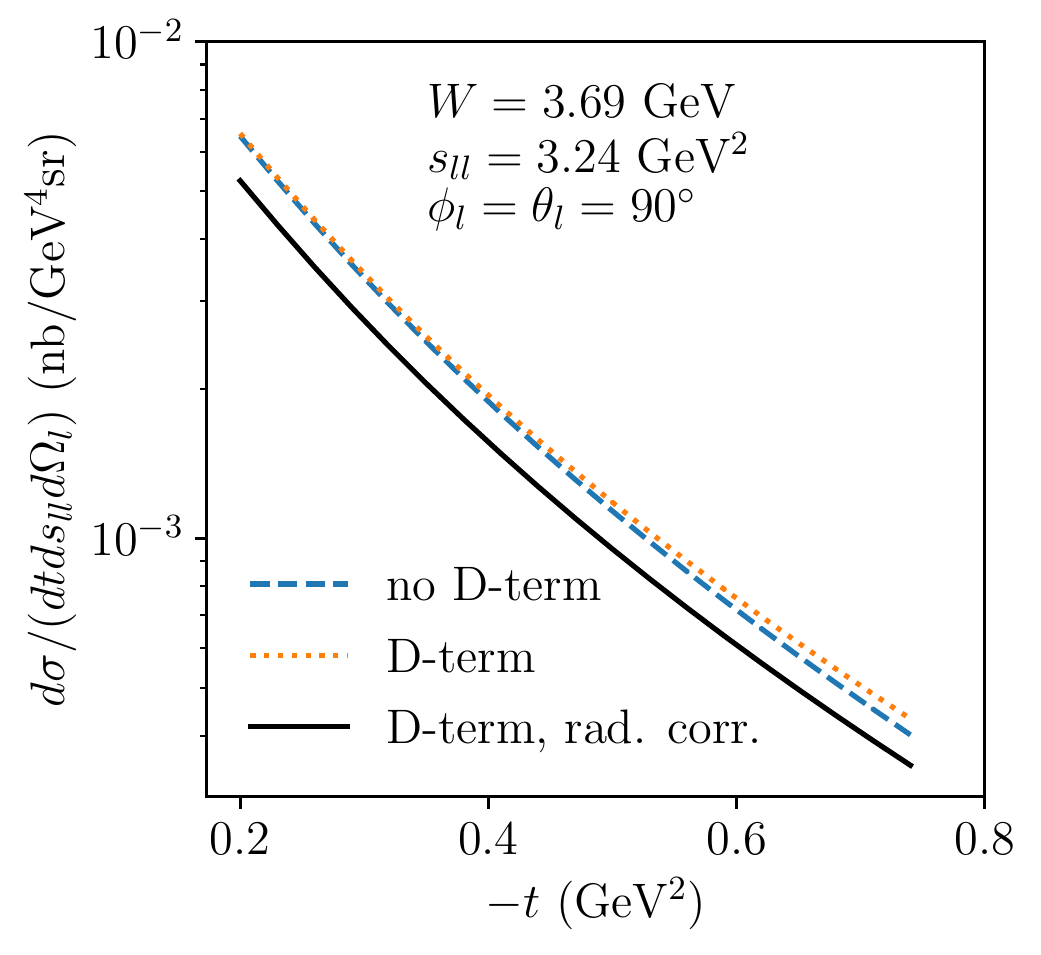}
 \includegraphics[scale=0.7]{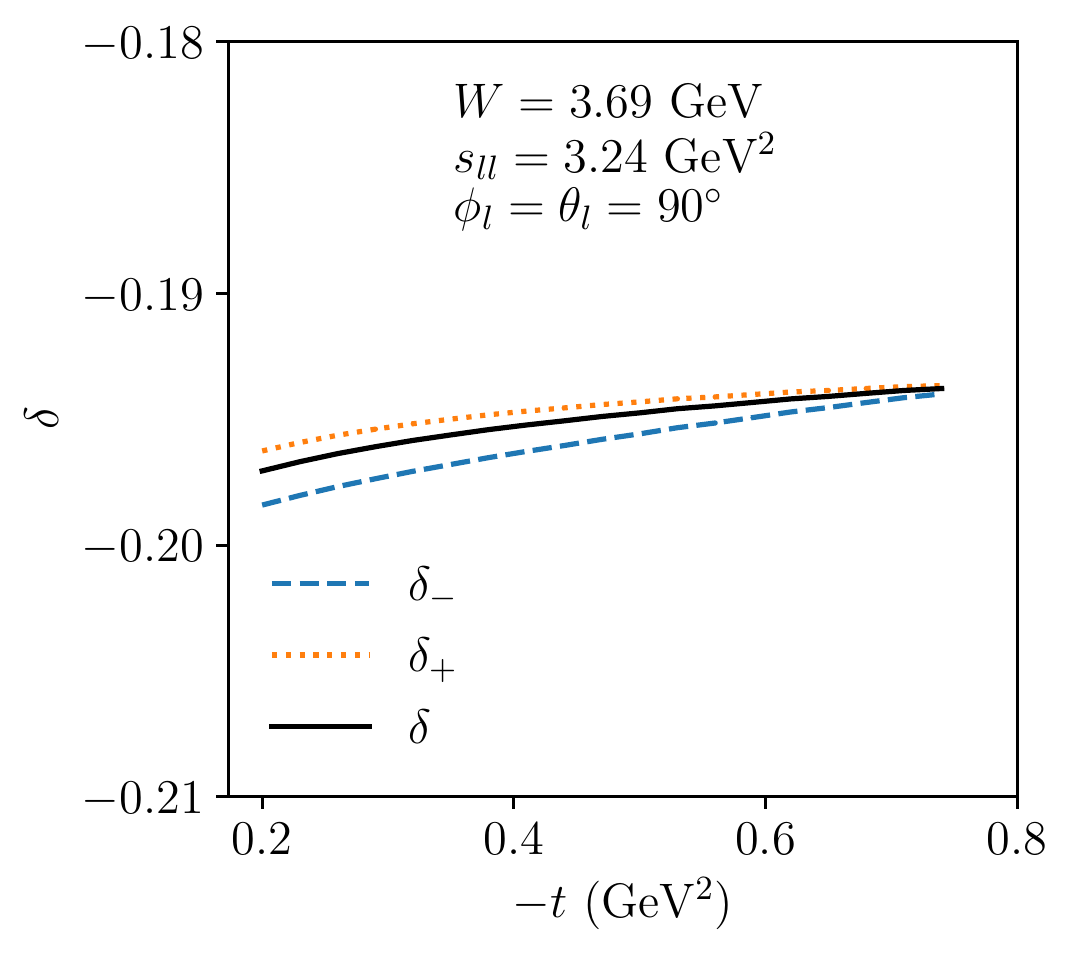}
  \includegraphics[scale=0.7]{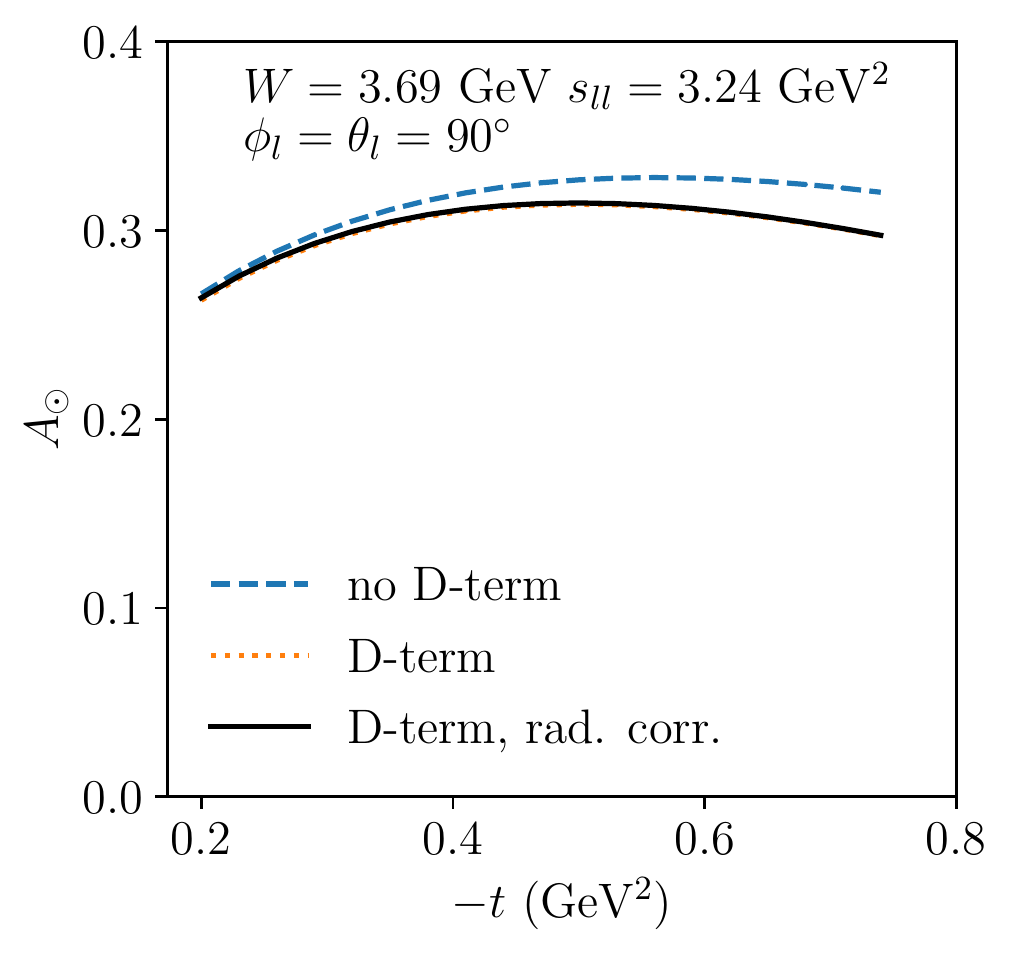}
	\caption{Upper panel: $t$-dependence of the $\gamma p \to e^-e^+ p$ unpolarized cross section for out-of-plane kinematics of the di-lepton pair corresponding with the CLAS12@JLab experiment~\cite{Chatagnon:2020egu,Chatagnon:2020}. We show our results for two GPD models and including the first-order QED radiative corrections, for a soft-photon energy of $\Delta E_s=0.05$ GeV. The middle panel shows the cross section correction factors $\delta_\pm$ for circular photon polarization $\pm 1$. The lower panel shows the sensitivity of the photon beam helicity asymmetry $A_{\odot}$ on the GPD parameterization, and the negligible radiative correction on this observable (solid and dotted curves nearly overlapping). \label{fig:high_energy_A_hel}}
\end{figure}

In Fig. \ref{fig:high_energy_A_hel} we show the 
$t$-dependence of the $\gamma p \to e^-e^+ p$ observables for a circularly polarized photon beam for out-of-plane di-lepton kinematics ($\theta_l = \phi_l = 90^\circ$), corresponding with the CLAS12@JLab experimental conditions. We notice that in out-of-plane kinematics the sensitivity to the D-term in the GPD parameterization is very small, while the radiative correction on the polarized cross sections is also around 20\%. In the corresponding photon beam helicity asymmetry $A_\odot$, which is directly proportional to the imaginary part of the BH-TCS interference, the first-order QED radiative corrections again nearly drop out. As the D-term is purely real, its effect on $A_\odot$ is very small, and contributes only through the squared modulus of the complex TCS amplitude. The imaginary part of the latter is rather well constrained by the precise electron beam spin asymmetry data for the corresponding DVCS process~\cite{Dupre:2016mai,Dupre:2017hfs}. As the non-perturbative part in both processes is the same, a direct comparison between the GPD predictions for DVCS and TCS beam helicity asymmetries at the same values of $-t$ and $\xi$, provides a stringent test of the applicability of the underlying QCD factorization theorem at these kinematics. Furthermore, as can be seen from the lower panel in 
Fig.~\ref{fig:low_energy_A_hel}, such test is basically unaffected by the first-order QED radiative corrections.

\section{Conclusions\label{sec:conclusion}}

In this paper we presented the first-order QED corrections on the lepton side contributing to the timelike Compton scattering process on a proton, $\gamma p \to l^-l^+ p$. 
This reaction contains contributions from both the Bethe-Heitler amplitude, for which the dependence on the proton structure is parameterized through its spacelike elastic form factors, and from the TCS amplitude with intial real photon and final timelike photon.   
We calculated the first-order radiative corrections on the level of the amplitude, such that the individual sub-processes can be easily incorporated into different calculations. 
We studied the effect of radiative corrections in two different energy regimes. In the $\Delta(1232)$ resonance region the TCS  amplitude was modeled as the sum of Born and $\Delta(1232)$-pole contributions, which was found to give a very good description of the 
near-forward kinematical regime. 
In the high-energy, near-forward regime we calculated the TCS amplitude in terms of GPDs, in kinematics of a recent CLAS12@JLab experiment. 

In the kinematics near the $\Delta(1232)$-resonance position, we found that the first-order QED radiative corrections on the $\gamma p \to e^-e^+ p$ cross section are in the 5 - 10\% range, while in the high-energy kinematics of the CLAS12@JLab experiment, they are in the 20\% range. Their inclusion is  thus important to extract the low-energy constants or Compton form factors from the $\gamma p \to e^-e^+ p$ process.  

Besides the corrections on the unpolarized cross section, we also studied in both kinematical regimes the effect of the first-order radiative corrections on the forward-backward asymmetry $A_{FB}$, obtained by interchanging the kinematics of the produced di-leptons, as well as on the photon beam helicity asymmetry $A_\odot$. While the asymmetry $A_{FB}$ accesses the real part of the BH-TCS interference, the asymmetry $A_{\odot}$ accesses the imaginary part of the BH-TCS interference in the regime where the BH amplitude dominates. The asymmetries $A_{FB}$ and $A_{\odot}$ are thus complementary in accessing experimentally the complex TCS amplitude. 

We found that in both kinematical regimes the radiative corrections on these asymmetries are at the few per mille level only, although the radiative corrections on the $\gamma p \to e^-e^+ p$  cross sections are in the 10 - 20\% range. The reason for the near cancellation of radiative corrections is that the corrections on the forward and backward cross sections as well as on both beam helicity cross sections are of the same size, and thus nearly cancel out in the corresponding ratios.
This makes the $A_{FB}$ and $A_\odot$ gold-plated  observables to extract the real and imaginary parts of the TCS amplitude respectively.   

In the $\Delta(1232)$ region, we find a value around +30\% for the forward-backward asymmetry, which provides a good opportunity for a direct measurement of the real part of the near-forward Compton amplitude. This will also allow for a comparison with existing dispersive extractions, in which the real part of the  forward Compton amplitude has been obtained as a dispersive integral over its imaginary part. At present the latter extraction is limited in precision due to  discrepancies in the world database for photoabsorption on a proton, especially in the $\Delta(1232)$ region, which also directly limits the obtained precision of the extracted Baldin sum rule value for the sum of proton electric and magnetic polarizabilities, $\alpha_{E1} + \beta_{M1}$. 

In the kinematical regime of the CLAS12@JLab experiment, 
the measurement of both asymmetries $A_{FB}$ and $A_\odot$ will allow to extract the Compton form factors from the TCS amplitude and compare them to the corresponding DVCS process with a spacelike initial photon and final real photon. As the non-perturbative part in both processes is the same, a direct comparison between the GPD predictions for DVCS and TCS observables, provides a stringent test of the applicability of the underlying QCD factorization theorem at these kinematics. 
In particular, the forward-backward asymmetry is a very sensitive observable to extract the D-term contribution to the GPD parameterization, as it only contributes to the real part of the TCS amplitude.
We found that a dispersive estimate of the D-term contribution leads to a shift in $A_{FB}$ by 10 - 20\%. The latter sensitivity will be interesting to test by the CLAS12@JLab TCS experiment, and is complementary to the information one extracts from the DVCS process, when comparing cross sections for both electron and positron beams. 

As a next step we plan to generalize our radiatively corrected calculations to the double virtual Compton scattering, in which the incoming photon has a non-zero spacelike virtuality while the outgoing photon has a non-zero timelike virtuality, accessed through the $e^- p \to e^- p l^-l^+$ reaction. In the $\Delta(1232)$ region, it was shown that this process allows for an empirical determination of the remaining unknown low-energy structure constant entering the hadronic correction to the muonic hydrogen Lamb shift ~\cite{Pauk:2020gjv}. In the high-energy near-forward region, double deeply virtual Compton scattering allows to extend the DVCS beam spin asymmetry measurements of GPDs into the ERBL domain~\cite{Guidal:2002kt,Belitsky:2002tf}. 
Since we calculated all sub-parts of the amplitudes for off-shell particles, it will be straightforward to generalize our setup to the double virtual Compton observables, accessed through the $e^- p \to e^- p l^-l^+$ reaction.

\section*{Acknowledgements}
This work was supported by the Deutsche Forschungsgemeinschaft (DFG, German Research Foundation), in part through the Collaborative Research Center [The Low-Energy Frontier of the Standard Model, Projektnummer 204404729 - SFB 1044], and in part through the Cluster of Excellence [Precision Physics, Fundamental Interactions, and Structure of Matter] (PRISMA$^+$ EXC 2118/1) within the German Excellence Strategy (Project ID 39083149).

\bibliography{biblio_tvcs}

\end{document}